%% file: main.tex
\documentclass[sigconf]{acmart}

\usepackage{graphics}      
\usepackage{xspace}
\def\BibTeX{{\rm B\kern-.05em{\sc i\kern-.025em b}\kern-.08emT\kern-.1667em\lower.7ex\hbox{E}\kern-.125emX}}

\newcommand{\topic}[1]{\vspace{-5pt}\smallskip \smallskip \noindent{\bf #1.}}

\newcommand{\automl}{Auto-ML\xspace}
\newcommand{\experts}{ML Innovators\xspace}
\newcommand{\intermediates}{ML Engineers\xspace}
\newcommand{\novices}{novices\xspace}

\setcopyright{rightsretained}

\begin{document}

\title{Whither AutoML? Understanding the Role of Automation in Machine Learning Workflows}

\author{Doris Xin}
\authornote{Equal Contribution}
\affiliation{%
  \institution{University of California, Berkeley}
}
\email{dorx@berkeley.edu}

\author{Eva Yiwei Wu}
\authornotemark[1]
\authornote{The majority of the work was performed when the author was at University of California, Berkeley}
\affiliation{%
  \institution{University of California, Berkeley}
  \institution{University of Zurich}
}
\email{eva.wu@berkeley.edu}

\author{Doris Jung-Lin Lee}
\affiliation{%
  \institution{University of California, Berkeley}
}
\email{dorislee@berkeley.edu}

\author{Niloufar Salehi}
\affiliation{%
  \institution{University of California, Berkeley}
}
\email{nsalehi@berkeley.edu}

\author{Aditya Parameswaran}
\affiliation{%
  \institution{University of California, Berkeley}
}
\email{adityagp@berkeley.edu}

\begin{abstract}
Efforts to make machine learning more widely accessible have led to a rapid increase in Auto-ML tools that aim to automate the process of training and deploying machine learning. To understand how Auto-ML tools are used in practice today, we performed a qualitative study with participants ranging from novice hobbyists to industry researchers who use Auto-ML tools. We present insights into the benefits and deficiencies of existing tools, as well as the respective roles of the human and automation in ML workflows. Finally, we discuss design implications for the future of Auto-ML tool development. We argue that instead of full automation being the ultimate goal of Auto-ML, designers of these tools should focus on supporting a partnership between the user and the Auto-ML tool. This means that a range of Auto-ML tools will need to be developed to support varying user goals such as simplicity, reproducibility, and reliability.
\end{abstract}


\copyrightyear{2021}
\acmYear{2021}
\acmConference[CHI '21]{CHI Conference on Human Factors in Computing
Systems}{May 8--13, 2021}{Yokohama, Japan}
\acmBooktitle{CHI Conference on Human Factors in Computing Systems
(CHI '21), May 8--13, 2021, Yokohama,
Japan}\acmDOI{10.1145/3411764.3445306}
\acmISBN{978-1-4503-8096-6/21/05}

\maketitle

\input{1_intro.tex}

\input{2_relatedwork.tex}

\input{3_studydesign.tex}

\input{4_results}

\input{5_discussions}

\section{Conclusion}
\label{sec:conc}

In this paper,
we presented results from a semi-structured 
interview study of 16 practitioners
who have used \automl in real-world applications.
Our participants
all reported various types of hybrid manual-auto
strategies in leveraging \automl in
their development workflow, providing
hints, safeguarding outputs, and massaging
inputs into a form digestible by \automl tools, among others.
Current work practices around \automl and perceptions of
these tools demonstrate that complete automation of ML
is neither realistic nor desirable.
Our study sheds light on various forms
of partnership or collaboration between humans and ML/AI,
depending on user motivations, needs, skill-set, and use-cases,
and provides next-generation \automl tool developers with
design guidelines for how to best incorporate pragmatic
guidance and empower effective engagement with \automl tools.

\smallskip
\noindent {\bf Acknowledgments.} We thank the anonymous reviewers for their valuable feedback. We acknowledge support from grants IIS-1940759 and IIS-1940757 awarded by the National Science Foundation, and funds from the Alfred P. Sloan Foundation, Facebook, Adobe, Toyota Research Institute, Google, and the Siebel Energy Institute. The content is solely the responsibility of the authors and does not necessarily represent the official views of the funding agencies and organizations.

\bibliography{refs}
\bibliographystyle{ACM-Reference-Format}

\end{document}

%% file: 1_intro.tex

\section{Introduction}

Machine Learning (ML) holds great promise
in providing solutions to some of our most pressing
societal problems, ranging from public health,
to climate change, to precision agriculture and food waste management.  
At the same time, ML is perceived to be a dark art, accessible
only to a few, involving
a cumbersome iterative process of trial-and-error
to craft a model with desired accuracy.
Users often have to experiment with a large set of modeling decisions, spanning data pre-processing and feature engineering, model types, and
hyperparameters, and empirically evaluate the performance of resulting models.

In recent years, automated machine learning, or \automl\footnote{Stylized as \automl to avoid confusion with Google AutoML.}, a new field targeted
at increasing automation during ML, has witnessed a rapid rise in popularity,
driving an explosion in self-proclaimed \automl tools. 
\automl holds the promise to make ML more easily accessible to users to employ for new domains and to reduce
cumbersome manual effort when applying ML to existing ones.
\automl was initially introduced to 
automate model hyperparameter search. 
Over time, \automl has evolved beyond that to include, 
as part of its goal, 
automation for other tasks in the ML workflow such as feature engineering, data cleaning, model interpretability, and model deployment. 
The ultimate promise of \automl tools 
is to make ML more accessible 
by providing off-the-shelf solutions 
for users with less technical backgrounds. 

In order for \automl tools to effectively meet user needs, we must first understand what those needs are and what roles automation currently plays in the ML workflow. Towards this goal, we conducted an in-depth, qualitative study of sixteen \automl users. We conducted semi-structured interviews with current users of \automl, ranging from hobbyists to industry researchers across a diverse set of domains, to take a careful look at their use cases and work practices, as well as their needs and wants. We asked about their uses of ML, their experiences working with and without \automl tools, and their perceptions of \automl tools. Our approach enabled us to not only gain 
a better understanding of user needs with respect to \automl, and 
the strengths and weaknesses of existing \automl tools, but also 
to gain insights into the respective roles of the 
human developer and automation in ML development.

Our findings indicate that human developers 
are indispensable in ML development. Not only do humans excel at bringing in valuable domain knowledge, human intuition can also be incredibly effective in filling in the blind-spots in \automl. Furthermore, human involvement is crucial for the effective and socially-responsible use of ML in real-world applications.
Therefore, rather than attempting to automate human developers ``out of the loop'' as has been the objective of many \automl tool builders, we advocate for a symbiotic relationship between the human developer and \automl to integrate the human into the loop in the most productive manner.
We, in fact, advocate that the moniker ``\automl'' be discarded, because
our evidence suggests that complete automation is infeasible;
instead, these tools can be better thought of as offering mixed-initiative ML solutions. 

Several prior studies have advocated for a human-guided approach to \automl~\cite{Lee2019AHP,gil2019towards,Dakuo2019Collaborative} and proposed design requirements. 
Rather than designing top-down, we build our case for human-AI integration based on bottom-up findings of real work practices, allowing us to arrive at unique, specific insights that are difficult to develop without taking the perspectives of the practitioner into account. 
We make concrete recommendations on what functionalities tool developers should enhance while preserving existing benefits, and, more importantly, what roles of humans they should preserve rather than attempt to replace. Our recommendations are based on both the perception and usage of a broad range of  users of state-of-the-art Auto-ML tools.
We hope that these insights and design recommendations can guide future development of \automl tools to expand the current focus on system and model performance to emphasize human agency and control.

The rest of the paper is structured as follows: we discuss the relationships between our work and relevant HCI contributions and present state-of-the-art on \automl tooling in Section~\ref{sec:rw}; we describe our study methodology in Section~\ref{sec:method}; we present findings on the benefits and deficiencies of current \automl tools as well as the respective roles of the human developer and automation in the ML workflow in Section~\ref{sec:results}; we discuss the design implications of our findings on \automl tools in Section~\ref{sec:discussions} and conclude in Section~\ref{sec:conc}.

%% file: 2_relatedwork.tex

\section{Related Work}
\label{sec:rw}
\automl systems are often aimed at developing an end-to-end ML workflow or model, unlike other human-in-the-loop ML tools that are more focused on specific parts of the ML workflow, such as collection of training examples, model debugging, and model interpretation. Our work builds on multiple areas of prior work: existing \automl systems, human-centered ML work practices, and human-in-the-loop ML tools. 

\subsection{\automl Systems}
\label{sec:tools}

The current landscape of \automl tools is fast-growing and diverse. 
Existing \automl offerings can be categorized into three groups: open-source software, hosted commercial solutions offered by cloud providers, and enterprise solutions offered by companies dedicated to developing \automl platforms. Cloud-hosted \automl solutions exist as part of a larger ecosystem of tools in the cloud, whereas enterprise solutions are standalone and therefore must either provide end-to-end support or integrate with external tools, resulting in appreciable differences in the user experience.

A typical ML workflow can be partitioned into three stages: \textit{data preprocessing}, \textit{modeling}, and \textit{post processing}, and \automl solutions provide varying levels of support for each of these stages. 
Below we offer a brief overview of tools in these three categories and comment on their support for the three stages in the ML workflow.

\topic{Open-Source Software (OSS)} 
\automl tools in this category include libraries such as Auto-sklearn (based on Scikit-learn)~\cite{NIPS2015_5872}, TPOT (based on Scikit-learn)~\cite{TPOT:Olson}, and AutoKeras (based on Keras)~\cite{jin2019auto}, all developed in academic labs, as well as TransmogrifAI (based on Apache Spark)~\cite{TransmogriFAI}, AdaNet (based on Tensorflow)~\cite{cortes2017adanet}, Ludwig~\cite{molino2019ludwig}, and H2O~\cite{H2O.ai} from industry. 
Of these tools, AutoKeras, AdaNet, and Ludwig are designed specifically for deep learning, tackling issues such as efficient neural architecture search, while the others are designed for traditional ML. 
Of the three categories of \automl tools, OSS tools are the best at keeping up with cutting-edge ML research since many of the OSS tool developers are also involved in ML research.
Overall, users of these libraries are afforded great flexibility since they can easily integrate custom code with the \automl API, but they must provision their own computation resources. 

For the specific stages in the ML workflow, the Scikit-learn-based libraries are better suited for structured data and have better support for automated data preprocessing than the other libraries, while the Tensorflow and Keras-based libraries can support more complex models involving text and images.
OSS tools are generally lacking on post-processing support, which involves evaluation, deployment, and monitoring of models. 

\topic{Cloud Provider Solutions}
The major players in this category are Google Cloud AutoML, Microsoft Azure Automated ML, and Amazon SageMaker Autopilot~\cite{GoogleCloudAutoML,MicrosoftAzureAutomatedML,AmazonSagemakerAutopilot}. 
Solutions in this category differ from the previous category in three significant ways:
1) Since they are hosted, compute resources are provided and managed by the cloud provider, and users pay proportional to the amount of compute consumed during the process of \automl; 
2) They tend to be much more end-to-end and include model evaluation and deployment to help users derive business value from the models trained;
3) Some system internals are opaque to the user, who can only interact with the system at specific decision points.
Overall, while cloud-hosted solutions tend to require less programming expertise to use, they are also \textit{less configurable and transparent}. 
For example, Google Cloud AutoML, which boasts ``more than 10 years of proprietary Google Research technology to help your ML models achieve faster performance and more accurate predictions'', neither allows users to specify the type of models nor provides visibility into the model internals. 
Amazon SageMaker Autopilot, on the other hand, places a strong emphasis on visibility and control by allowing users to easily export the \automl code and intermediate results into computational notebooks. 
All three providers offer no-code UIs for non-programming users, in conjunction with Python APIs.
While Microsoft and Amazon enable additional customizability through their Python APIs, Google's APIs are solely designed for programmatic compatibility and offer no additional control or transparency.
When it comes to model evaluation, Microsoft and Amazon provide summaries of the models explored while Google offers only high-level information about the final model.

\topic{\automl Platforms}
As self-proclaimed ``platforms'', tools in this category tend to position themselves as turnkey, end-to-end \automl solutions. 
They manage compute resource provisioning by integrating with either cloud providers or on-premise hardware infrastructure.

Major players in this category include DataRobot~\cite{Datarobot} and H20 Driverless AI~\cite{H2O.ai}. 
 
Compared to their cloud provider counterparts, solutions in this category tend to be more feature-complete, providing more technical support and customizability in each stage of the workflow. 
Since these solutions target business users, special attention is paid to the operationalization of the resulting model, including an expansive set of model interpretability and deployment options.
Additionally, to address increasing concerns around data privacy and security, these solutions allow on-premise deployment instead of forcing users to migrate their data to the cloud. 

\vspace{3pt}
Thus far, there has been little evaluation on how these three types of \automl tools are used in practice. In this paper, we sought to understand the adoption of \automl tools, their usage, and the current bottlenecks that users face with these tools. Our eventual goal is to identify design requirements and considerations for more effective collaboration between the human developer and ML/AI.

\subsection {Human-Centered ML Work Practices} 
HCI research in \automl has proposed a human-guided approach to ML~\cite{gil2019towards,Lee2019AHP,Dakuo2019Collaborative}, with the aim of balancing the trade-off between user control and automation. Human-guided ML builds on the premise that ML development should be a collaborative activity between human users (i.e., data scientists or model developers) and the machine, wherein users specify their domain knowledge or desirable model behavior at a high-level and the system performs some form of automated search to generate an appropriate ML pipeline or model~\cite{gil2019towards,Lee2019AHP}. 
This work is related to a larger body of studies on ML work practices~\cite{Kandel,Amershi2019Software,YangNonExpert,hong2020human,hohman2020understanding} of specific groups of practitioners, including software engineers using ML~\cite{Amershi2019Software}, non-expert ML users~\cite{YangNonExpert}, as well as specific aspects of work practices, such as model interpretation~\cite{hong2020human} and iterative behavior in ML development~\cite{hohman2020understanding,Lee2020}. \par Existing work has explored the use of visualization to help users gain insights into the black-box process of \automl and obtain higher control during \automl~\cite{wang2019atmseer, weidele2019autoaiviz}.
Another crucial step towards human-guided ML requires understanding the perspective of data science practitioners' and their attitudes towards \automl systems~\cite{Dakuo2019Collaborative}.
In particular, 
Drozdal et al. study the perceptions of transparency and trust in Auto-ML and identify components within Auto-ML tools most likely to improve trust~\cite{Drozdal2020TrustAutoML}. We expand upon this work by studying the role of humans and Auto-ML in the end-to-end ML workflow, as well as additional important user requirements, such as customizability, completeness, ease-of-use, efficiency, effectiveness, and generalizability.

These prior studies have largely focused on participants without \automl experience. As a result, most of these studies elicited user feedback regarding \automl by demoing new \automl prototypes to study subjects. These prototypes have limited features, largely supporting only the modeling stage of the ML process. 
For example, the study by Drozdal et al. involves university students and uses hypothetical tasks to determine the relative value of components of \automl in communicating trust, rather than centering on the real-world experiences of \automl tool users.
Our paper builds on this line of work by studying users who have worked with present-day \automl tools in real-world applications to investigate how \automl fits into their end-to-end ML workflow, as well as its limitations. 
Our methodology and choice of participants afford novel findings beyond existing studies on the matter of trust and transparency, such as the broader social context surrounding the use of \automl in the real world, e.g.,  getting others in the organization to adopt the results from an \automl model, or \automl playing a role in reproducibility and institutional knowledge should the original user leave the organization. Moreover, we identify various ways hands-on human agency and control promotes trust---something left unaddressed by studies involving hypothetical \automl interfaces.
 
\subsection{Human-in-the-loop ML Tools}
The idea of incorporating human knowledge into ML workflow development has been well studied in research on interactive tools for ML model interpretation. Interactive machine learning (IML) is a paradigm wherein human users train an ML model by manually evaluating and correcting the model result through a tight interactive feedback loop~\cite{Fails2003,Fiebrink2011,ramos2020interactive,Amershi2014}. IML often focuses on collecting user input for labels for training data in order to train and correct ML classification models. IML lowers the barrier to interacting with ML-powered systems by empowering end-users without ML expertise to interactively provide exemplar feedback to the system ~\cite{Amershi2014,YangNonExpert} (e.g., recommendation systems can learn from the relevance feedback of approving/rejecting an item), often eliminating the need for the highly-technical feature engineering step~\cite{Fails2003}. In addition to IML, human-in-the-loop systems have also been developed for model debugging and verification~\cite{Patel2010,amershi2015modeltracker, ono2020pipelineprofiler}, iterative model development~\cite{xin2018helix}, and model interpretation~\cite{mlfacets,Wongsuphasawat2018,Talbot2009} often through an interactive environment and/or visualizations. We study the impact of such capabilities as reflected in present-day \automl tools. 

Prior work questions the validity of humans-in-the-loop~\cite{AgainstHITL2019}. However, the study experiments only examined the effects of humans as the final arbitrators, anchoring on ML recommendations. Our study shows that humans are in the loop throughout the ML workflow even as \automl attempts to automate away ML development. Along the path, humans can iteratively influence the directions of ML outputs via several touch points. Further research is necessary to understand the effects of mixing ML and humans in real-world settings.

%% file: 3_studydesign.tex

\section{Study Design}
\label{sec:method}
Our paper seeks to understand how users incorporate \automl tools in their existing workflows and their perceptions of the tools. 

\subsection{Recruitment}
Since, to the best of our knowledge, no prior work has focused on participants with real-world \automl experience, we focused on users who have applied \automl to real-world use cases across a broad set of application domains. We recruited participants by posting recruitment messages to relevant mailing lists and Slack channels (N=5), through personal connections (N=9), and by reaching out to users on social media who mentioned or were mentioned in posts about \automl (N=2). 

We invited participants to fill out a screening survey that asked whether they had experience using \automl tools. Select participants either had experience using at least one of the tools that were listed in the survey (the list includes tools discussed in Section~\ref{sec:tools}) or had used other tools that our team verified to be \automl tools. The tool-based eligibility criteria was motivated by our pilot recruitment strategy where a large number of survey respondents expressed that they had general ML experience but were unfamiliar with \automl, mistaking manual ML for ``automated ML''.

\subsection{Participants}
 We interviewed a total of 16 participants who had prior experience applying \automl in professional capacities. Information about each participant is shown in Table~\ref{tab:participantdescriptors}. Of the 16 participants, 14 were male (87.5\%) and 2 were female (12.5\%). Recruiting participants past March 2020 was difficult due to COVID-19, but we gathered a sample of users that spanned a diverse set of organizations and use cases.

 Participants had, on average, 10 years of experience in programming and an average of 5 years of experience with ML.
 Participants spanned across three continents and a diverse set of job roles and industries, from a product manager at a large retail corporation, to a director of commercial data science at a travel technology company, to academic researchers at universities. The tools that the participants used also varied from proprietary tools, to open-source software, to commercial solutions. To preserve participant anonymity, we omit the identities of the specific tools used, as some tool-application combinations can reveal the identity of the participant.

\begin{table*}
  \caption{Interviewee demographics, from the left: (1) Participant ID, 2) Participant's years of experience (Yrs of Exp) in programming and in ML, (3) Domain of their companies and organization size in parentheses with ``(S)'' corresponding to small, ``(M)'' to medium, and ``(L)'' to large, (4) Participant job title, (5) \automl application, (6) Type of \automl tool(s) used by the participant. ``(CP)'' indicates that the commercial solution is provided by a major cloud provider and (PF) indicates that the commercial solution belongs to the \automl platform category introduced in Section~\ref{sec:tools}.
  }
  \label{tab:participantdescriptors}
  \small
  \begin{tabular}{p{0.5cm}p{1.8cm}p{2.5cm}p{2.8cm}p{3.5cm}p{2cm}}
    \toprule
    PID & Yrs of Exp (programming/ML) & Organization Domain (Organization Size) & Participant Role & Application & \automl Tool Category\\
    \midrule
    \textbf{P1}& 30/5 & Finance (M) & Data Science & Financial credit scoring & Commercial (PF)\\
    
    \hline
    \textbf{P2} & 8/2 & Healthcare (S) & Data Science & Object identification in videos & Commercial (CP) \\
    
    \hline
    \textbf{P3} & 4/7 & Retail (L) & Product Management & Fraud detection in e-commerce transactions  & Commercial (PF)\\
    
    \hline
    \textbf{P4} & 11/6 & Social Network (L) & ML Engineering& Content curation and recommendation & Proprietary\\
    
    \hline
    \textbf{P5} & 10/3 & University (L) & Biomedical Research & Drug response prediction & Open Source\\
    \hline
    \textbf{P6} & 10/4 & Healthcare (S) & Data Science & Medical claims analysis for healthcare program referral & Commercial (PF)\\
    
    \hline
    \textbf{P7} & 12/2.5 & Finance (L) & Data Science & Fraud detection in financial transactions & Commercial (CP)\\
    
    \hline
    \textbf{P8} & 5/4 & University (L) & Neuroscience Research & 
   
    Diagnostics and phenotype analysis with brain scan images
    & Open Source\\
    
    \hline
    \textbf{P9} & 8/8 & ML Software (S) & CTO & 
    
    Enterprise document processing automation
    & Commercial (CP)\\
    
    \hline
    \textbf{P10} & 20/8 & Travel Technology (M) & Data Science & Personalized travel recommendations & Commercial (PF)\\
    
    \hline
    \textbf{P11} & 6/5 & Information System (S) & ML Research & 
     
    Computer vision for brand logo detection
    & Open Source\\
    
    \hline
    \textbf{P12} & 16/12.5 & Educational Software (M) & Data Science & 
    Content identification in educational materials  
    & Commercial (CP)\\
    
    \hline
    \textbf{P13} & 6/3 & Electronic Product Manufacturing (L) & Research & Gesture recognition 
    in sensor data from edge devices & Commercial (CP)\\
    
    \hline
    \textbf{P14} & 12/5 & Consulting (S) & Software Engineer & Nonprofit AI consulting for businesses & Open Source\\
    
    \hline
    \textbf{P15} & 4/3 & Pharmaceutical (L) & ML Researcher & 

    Drug response prediction
    & Open Source\\
    
    \hline
    \textbf{P16} & 2/1.5& Supply Chain (L) & Program Management & Supply chain procurement and risk-management& Proprietary\\

  \bottomrule
\end{tabular}
\end{table*}

\subsection{Interview Procedure}

We conducted semi-structured interviews with participants about their experience in using \automl for real-world applications. Interviews were conducted from October 2019 to March 2020, either in person (N =2) or remotely (N=14). Each interview lasted for approximately one hour. Every participant received a $\$15$ gift card for compensation.
The interviews were largely semi-structured and involved three main components: 
\begin{itemize}
    \item Participants were asked to describe their job role, organization, and ML use cases.
    \item We asked the participants about their experience in developing ML workflows (without \automl) and the challenges they faced.
    \item We asked the participants about their experience using \automl tools, including the features of the specific tool used and how they integrated \automl in their ML workflows, and their perceptions of \automl tools in general, including customizability, effectiveness, interpretability, and transparency.
\end{itemize}
 \noindent The interview guide can be found in the supplementary materials.

\subsection{Study Analysis}
We audio-recorded and transcribed all but one interview, where the participant did not consent to being recorded. After completing all the interviews, we engaged in an iterative and collaborative process of inductive coding to extract common themes that repeatedly arose in our data. Each of the first three co-authors independently coded the data using Dedoose~\cite{dedoose}, an online tool for open coding, to map data onto these categories. 
The first three authors met weekly and discussed themes and concepts to clarify ambiguity in the codes and established consensus in a code book.
Afterwards, we conducted a categorization exercise, wherein some of our initial categories included advantages and disadvantages of \automl, perceptions of \automl, workflow strategies with and without \automl, \automl desiderata, general ML challenges, and \automl adoption decisions. We used codes to facilitate the process of theory development and refrained from calculating inter-rater reliability to avoid potential marginalization of perspectives~\cite{McDonald2019InterraterReliability}.

\subsection{Limitations}

Our interviews were structured around the participants' experience with a single ML task. While this practice afforded concreteness to the discussion, we may have missed out on diversity of use cases. 
The interviews focused on work practices around \automl, which allow us to gain a deep understanding of the strategies participants employed to integrate \automl into their data science workflow, but limited our time in studying any particular conceptual perception (i.e. transparency, interpretability) of \automl in-depth. 

We acknowledge that while our participant population comes from diverse background, it may not be a representative sample of the overall \automl user base. 
For example, the proportion of women in our sample (14 male, 2 female) is slightly lower than the gender ratio (15\% women) in the data science profession referenced in the 2020 BCG report~\cite{GenderRatio}.

%% file: 4_results.tex
\section{Results}
\label{sec:results}

In this section, we present our findings from the interviews. This section is organized as follows: we describe how we group the users and use cases into high-level categories in Section~\ref{sec:segment}; we enumerate common tasks in ML workflows reported by the participants in Section~\ref{sec:tasks}; we present the benefits of \automl perceived by the participants in Section~\ref{sec:benefits}, the deficiencies of existing \automl tools that we believe can be addressed by system and UI improvements in Section~\ref{sec:deficiencies}, and the roles of the human developer that the \automl tools must respect and preserve in Section~\ref{sec:humanRole}. 
Note that while the functionalities presented in Section~\ref{sec:deficiencies} and Section~\ref{sec:humanRole} are both absent or lacking in existing \automl tools, the deficiencies as described in Section~\ref{sec:deficiencies} have near-term solutions using existing techniques (we provide many such suggestions in Section~\ref{sec:discussions}), whereas the roles in Section~\ref{sec:humanRole} may one day become within reach of \automl tools through fundamental breakthroughs in knowledge representation and programming paradigms. We believe that assuming the roles in Section~\ref{sec:humanRole} should not be the objective of \automl tool developers until then.

\subsection{User and Use Cases Segmentation}
\label{sec:segment}
A given participant's ML skill-set and use cases can potentially influence how they use and perceive \automl tools. We, therefore, wanted to study the relationship between users' expertise and their behavior around and perception of \automl. To facilitate the discussion on how contextual information serves to explain user behavior and sentiment, we categorize the participants and their use cases as follows. 

\subsubsection{User Skill Levels}\hfill\\
We group participants based on their past experience with ML into the following three categories.

\topic{ML Innovators}
Participants in this group have formally conducted ML research, either on fundamental algorithms (P4, P11, P14, P15) or the application of ML for scientific discoveries (P5, P8). They command deep understanding of the mathematical underpinnings of the ML models they use.

\topic{\intermediates}
Participants in this group are skilled and experienced ML practitioners with formal training in applied ML (P1, P2, P3, P6, P7, P10, P12, P13). They are intimately familiar with popular ML models as well as ML libraries and tools. 

\topic{Novices}
Participants in this group (P9, P16) have no formal training in ML. They are domain experts in non-ML fields.

\subsubsection{Use Case Categories}\hfill\\
\topic{Production Applications}
Use cases in this category pertain to developing ML models that drive applications or decision making with significant impact. Examples include training models to assist in financial service decisions, building recommendation systems for content curation, and fraud detection.
P1, P3, P4, P7, P12, and P16 have use cases in this category.

\topic{Prototype}
Use cases in this category involve prototyping ML models for industry applications, which are lower stake than the production application scenario since the model performance has no direct impact on business metrics or end-user experience. 
P2, P6, P9, P10, P11, P13, P14, and P15 have use cases in this category.

\topic{Research}
Use cases in this category involve building ML modelings for academic research, where model results are shared and examined in detail.
P5 and P8 build ML models for research.

\begin{figure*}
    \centering
    \includegraphics[width=0.9\textwidth]{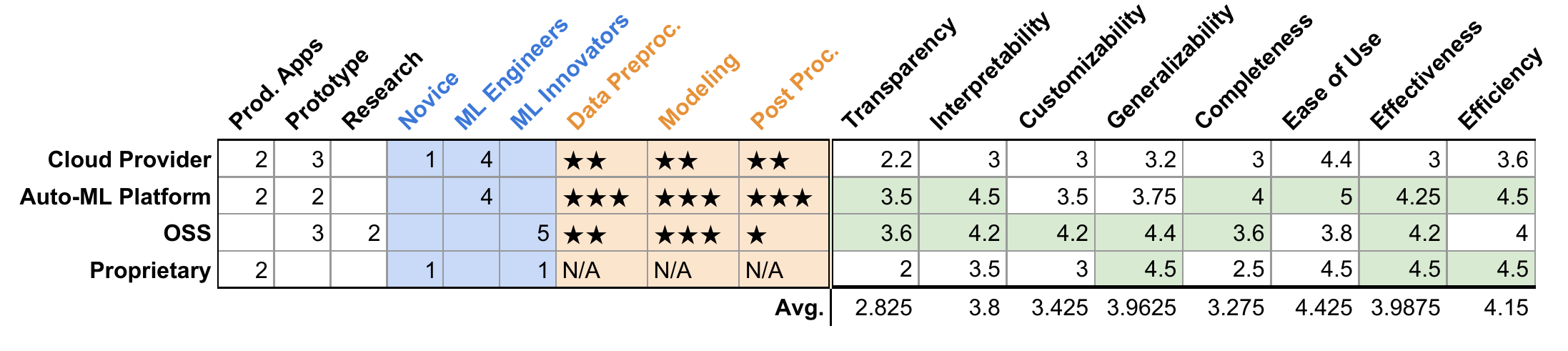}
    \caption{Tools: Characterization of \automl tools used by participants by category. The tool categories are the same as the ones presented in Section~\ref{sec:tools}, with the addition of ``proprietary'' for anonymous in-house \automl solutions.
    The first three columns contain the cross tabulation of tool categories and use case categories. The next three columns in blue contain the cross tabulation of the tool categories and user expertise levels. The following three orange columns indicate the level of support for each of the three ML workflow stages based on features offered (not based on interview results). The next eight columns contain the average Likert scores provided by our participants. Cells in green are above average.}
    \label{fig:tools}
\end{figure*}

Figure~\ref{fig:tools} shows the cross tabulation of the \automl tools category with use case categories and user expertise, the level of support for each of the three ML workflow stages based on features offered by each category of tools, and the average Likert scores that our participants gave to their \automl tools.
The support ratings for ``proprietary'' tools are omitted as we do not have full visibility into the these tools. While Likert scores are commonly treated as ordinal data, they are treated as interval data in our setting due to the natural correspondence between the scores and quintiles. Thus, the average Likert scores are sound and meaningful.

\subsection{Data Pre-processing, Modeling, and Post Processing Tasks}
\label{sec:tasks}
\input{4_s2_use_cases}

\subsection{Benefits of \automl}
\label{sec:benefits}
\input{4_s3_benefits}

\subsection{Deficiencies of \automl}
\label{sec:deficiencies}
\input{4_s4_deficiencies}

\subsection{Roles of the Human in \automl}
\label{sec:humanRole}
\input{4_s5RoleHuman}

%% file: 4_s2_use_cases.tex
\begin{figure*}
    \centering
    \includegraphics[width=0.9\textwidth]{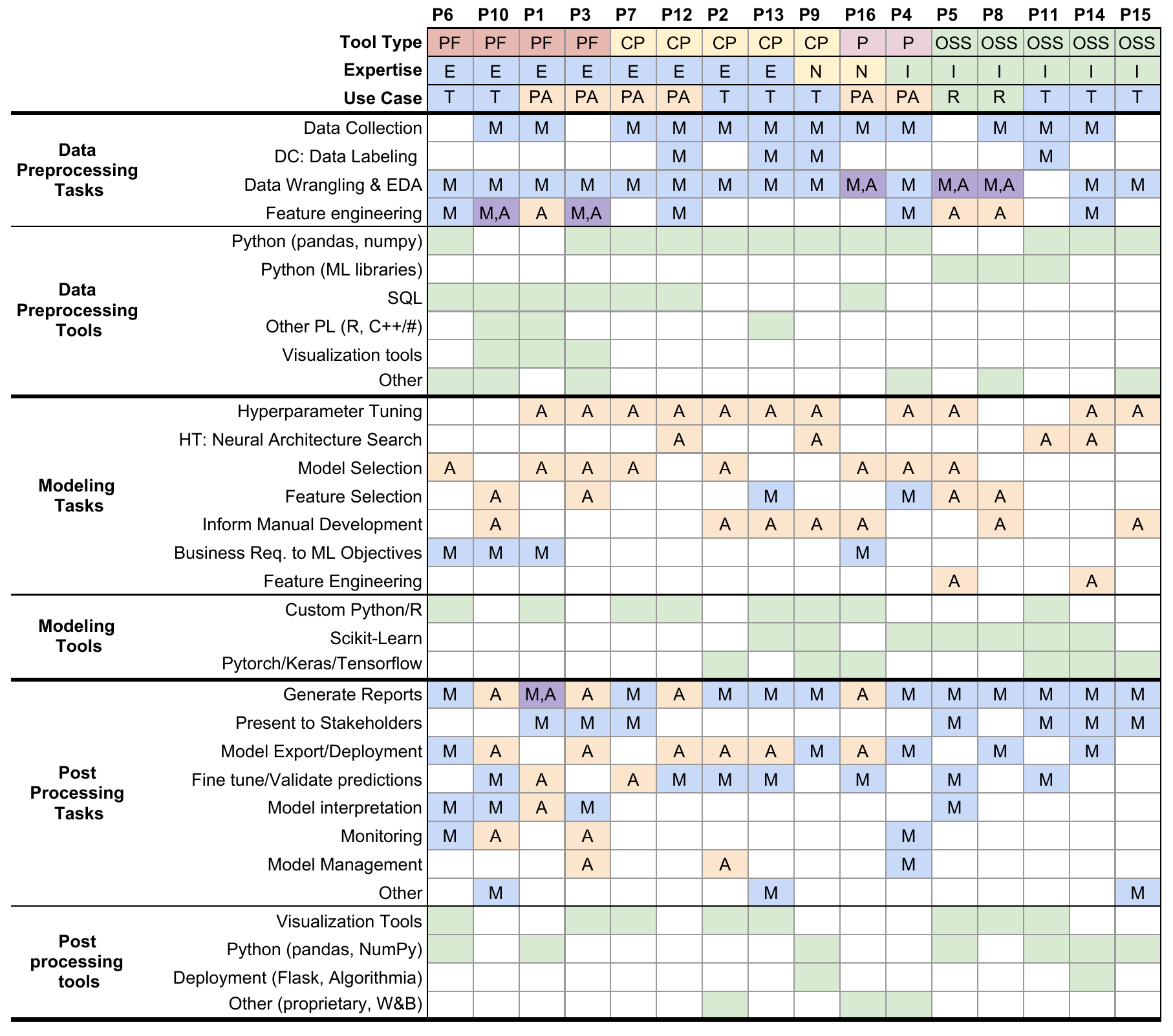}
    \caption{Use cases: tasks performed and tools used by participants in each stage of the ML workflow.
    The three rows at the top contain metadata about each participants. 
    ``Tool type'' corresponds to the tool types in Figure~\ref{fig:tools}: PF = \automl platform, CP = Cloud Provider, and P = Proprietary.
    ``Use case'' corresponds to the use case categories introduced in Section~\ref{sec:segment}: T = Prototype, PA = Production Application, R = Research.
    ``Expertise'' corresponds to the user skill levels introduced in Section~\ref{sec:segment}: E = \intermediates, N = \novices, I = \experts.
    In each task cell, ``M'' indicates that the participant performed the task manually while ``A'' indicate automation. A green tool cell indicates that the participant has used the tool for the given ML workflow stage.
    Columns as ordered to cluster participants who use the same type of tools and have the same expertise levels.
    }
    \label{fig:use_cases}
\end{figure*}

In this section, we focus on how practitioners integrate \automl tools into the end-to-end ML workflow. As mentioned in Section~\ref{sec:tools}, an ML workflow is commonly partitioned into three stages: data pre-processing, modeling, and post-processing. In this section, we report the common patterns of how \automl fits into the each stage.  Figure~\ref{fig:use_cases} shows a complete set of tasks the participants reported for each stage of the ML workflow. 

We characterize these three stages as follows:
\begin{enumerate}
    \item In the data pre-processing stage, users prepare the data for ML, performing tasks including data acquisition, cleaning, transformation, labeling, and feature engineering. 
    \item In the modeling stage, models are trained on the data prepared in the pre-processing stage.
    \item After models are trained, the post-processing stage span a broad range of activities, including model evaluation, interpretation, deployment, and user studies.
\end{enumerate}

\subsubsection{Data Pre-processing}
\label{sec:useCaseDPR}
Overall, data pre-processing is a time consuming and primarily manual
process for most participants. 
On average, participants reported spending roughly 50\% of their time on
data pre-processing, and roughly 80\% of all data pre-processing tasks were performed manually.
While some participants expressed the 
desire for Auto-ML to provide more automated support for data
pre-processing, others 
believe that data pre-processing relies on human intuition and knowledge
that is impossible or at least 
extremely challenging to codify and therefore cannot be automated.
In either case, participants agree that data pre-processing is a crucial component in the workflow
due to the adage ``garbage in, garbage out.'' We defer discussions on desired but missing automated
data pre-processing features and the role(s) of the human to Section~\ref{sec:deficiencies} and Section~\ref{sec:humanRole}, respectively.
Here, we report on the common tasks in the 
data pre-processing stage and whether each task is
currently carried out by Auto-ML or human developers.

\topic{Data collection} 
The first task in data pre-processing is 
\textit{data collection}~\cite{roh2019survey} for most participants except P3, P5, P6, and P15, who were provided and limited to specific data sources.
Others had the freedom to incorporate organizational data assets obtained from relevant stakeholders (P4, P7, P8, P9, P10, P11, P12, P13, P14, P16) as well as open datasets (P1, P2) from the web, which are much less commonly used. 

A specific data collection task worth noting 
is \textit{data labeling}, where significant 
human attention is required to annotate examples with the desired target label. 
P9 and P13 reported performing labeling themselves because it provided them with intuition about the data and allowed for fast turnaround. P11 and P12, both of whom worked with unstructured data, relied on crowdsourcing for data labeling. 
Data labeling is inherently manual, although there are recent developments in systems to lower the manual effort for data labeling~\cite{bach2017snorkel,bringer2019osprey}.

\topic{Data Wrangling \& EDA}
Following data collection is data wrangling, 
including data cleaning and missing data imputation, formatting, and transformation. 
Exploratory data analysis (EDA) is integral to all data wrangling tasks. 

All participants except P11, 
who worked with image data, reported that they spent time on data wrangling.
In addition to \automl tools used by P5, P8, P16 to help with data wrangling, the Python pandas~\cite{mckinney-proc-scipy-2010} and numpy~\cite{van2011numpy} libraries are the most popular choices for data wrangling, followed by SQL. 

Tools in the ``Other'' category in Figure~\ref{fig:use_cases} 
include domain specific tools, Spark, and proprietary tools.

Data cleaning, especially missing data 
identification and imputation, is a common task.

While most \automl tools handle missing data imputation, participants stated manual intervention is necessary to decide on the appropriate imputation technique based on context (P1, P4, P5, P7).
Participants reported that they spent significant manual effort to format the data for ingestion by the \automl tool (P2, P3, P8, P10, and P12.) 
Specialized 
data wrangling tasks included data anonymization (P3), time series test set generation (P10), and resampling for class imabalance (P16).

\topic{Feature engineering}
Feature engineering and feature selection 
are among the most automated data pre-processing tasks. 
Most \automl tools are capable of both simple feature engineering 
such as one-hot encoding~\cite{wiki:one-hot} 
and building complex features 
involving multiple input signals. 
However, over half the participants who performed feature 
engineering manually because they felt 
existing \automl solutions 
are incomplete or inefficient (P6, P10, P14) or
they believed that 
human intuition and 
domain knowledge could not be replaced by automation (P3, P4, P12). Interestingly, P5 and P14 both reported repurposing, or \textit{``misusing''} (P14) modeling capabilities for feature engineering.

\subsubsection{Modeling}
As one would expect, the most common tasks that participants used \automl for during modeling are 
\textit{hyperparameter tuning} and \textit{model selection}. 
\textit{Feature selection} is generally carried out by the \automl tool as a byproduct of model training. Participants 
who perform feature selection 
all use \automl for feature selection except P4, 
who manually selects features 
before model training because 
they work with petabyte-scale data, and P13,  
who deploy models to edge devices with limited memory.

An interesting use case for \automl modeling 
is to \textit{inform manual development} 
(P2, P8, P9, P10, P11, P13, P15.) 
For example, P13 reported that 
they use the \automl tool as a quick 
check for data quality and 
to validate manual data pre-processing. 
They would perform data wrangling manually if the \automl tool 
identified anomalies in the data.
\automl also empowers \textit{exploration of unfamiliar models and hyperparameters} (P2, P8, P10, P13, P15, and P16.)

\automl results are often used 
as a \textit{benchmark for validating 
manual model performance} (P4, P5, P9.)

Many participants reported that they performed the same modeling tasks manually  
alongside \automl to understand and verify the \automl results and to correct any errors made by \automl.

The modeling tools in Figure~\ref{fig:use_cases} refer to tools for 
manual modeling.

\subsubsection{Post-processing}
Post-processing spans a large variety 
of tasks depending on 
the participants' use cases. 
Visualization is an integral part of many of the tasks below. 

\topic{Generating reports and sharing results}
The most common post-processing task 
is to generate a report of the 
model results and relevant model search history. 
Most platform tools automatically generate 
such reports, 
in the form of summary statistics, leaderboards, 
and other visualizations. 
P10 enthusiastically shared that 
their platform tool was able to 
auto-generate documentation 
for legal compliance thereby 
greatly reducing 
the manual overhead for governance. 
While all of the 
cloud-hosted \automl tools also auto-generate reports and 
visualizations, it is interesting that many participants adopted 
manual approaches to modify the default reports. 
In addition to generating reports 
for their own consumption, a subset of the participants (P1, P3, P5, P7, P11, P14, P15) who had to share and explain their results to 
other stakeholders needed to present their findings in text documents or slides with human readable explanations. 

\topic{Deploying Models}
Model export and deployment is the second most common 
post-processing task.
Participants who did not 
perform model deployment were either using the models to inform 
human decision making (P1), 
handed off the model to a separate 
Dev Ops team for deployment (P7, P11) or used model results solely 
for research findings (P5, P15). 
Automated deployment was 
only afforded to users of hosted \automl tools. 
The reason for manual 
deployment for participants 
who used hosted tools include 1) the model for a financial application needed to be vetting for security (P7), 
2) the \automl tool did not support automated deployment (P6), 
3) the application relied on complex logic to incorporate the model output (P9), and 
4) the model directly impacted end-user experienced and required staggered roll-out with human supervision (P4).
Tools for deployment included Flask, Algorithmia, 
custom Python, and proprietary infrastructure. 

\topic{Validating and Interpreting Models and Predictions}
The two main techniques for building trust in and understanding \automl outputs are point queries on the predictions, feature importance, and holistic visualization of high-level model characteristics. 
Most hosted solutions focus on supporting point queries and feature importance via visualizations. Manual efforts in this category were mainly for 
1) cross validating with manual modeling results (P3, P10, P11),
2) application domain specific checks (P6, P12),
3) field testing specific predictions (P13, P16),
4) custom feature importance computation (P5).

\topic{Miscellaneous}
Less common, nevertheless noteworthy tasks included model versioning (P2, P3, P4), on-device user studies (P13, P15), debugging manual implementation (P15), and most interestingly, converting an ML model into a set of if-else statements for more predictable and interpretable inference (P10).

%% file: 4_s3_benefits.tex

In this section, we present findings on the major benefits of \automl from our interview study. As shown in Figure~\ref{fig:tools}, ease of use, effectiveness, and efficiency are the highest-rated qualities of \automl tools by the interview participants. We present specific benefits reported by the participants that corroborate these ratings below.

\topic{Enables and empowers \novices} 
To ML \novices, the greatest benefit of \automl is that it enables business users to nimbly use 
ML to inform business decisions without ``\textit{a massive engagement with multiple consultants in multiple different teams}''(P16). 
P16 believes that \automl leads to ``\textit{the democratization of advanced analytics throughout business units for people that don't have experience doing that kind of work}'', and this sentiment is echoed by P10:
\begin{quote}
    ``It allows for \ldots citizen data science to become a reality with the proper governance controls and proper management in place. At the bank I worked at previously \ldots \automl was becoming adopted \ldots for robotic process automation. \ldots Anyone who's analytically competent \ldots starts rolling with it immediately.''
\end{quote}

However, \automl can be a double-edged sword 
for novice users---users who treat it 
as largely a blackbox find \automl 
to be a great enabler, while curious 
users who attempt to look inside the blackbox can become distracted and suffer from choice paralysis. 
P9 reported that while they were able to achieve a few percentage points in model performance improvement, \automl \textit{increased} their development time, due to the fact that 
the \automl tool exposed them to a large number of new model types that led to many lengthy manual explorations out of curiosity.

In addition to avoiding distractions, 
treating \automl as a blackbox also leads to standardization.

\topic{Standardizes the ML workflow for better reproducibility, code maintainability, knowledge sharing}
Another benefit of the blackbox nature of \automl tools is that
by having a predetermined search space that doesn't change,
there is more 
standardization of the ML development process, leading  
to better comparisons 
across models, code maintainability, and effortless knowledge transfer. 
The need to search through a large number of model types, which are often implemented in different libraries, has prompted \automl tools to create a standardized abstraction of the ML workflow decoupled from specific APIs and model types. As a result, 
\begin{itemize}
    \item different models are easily comparable using standardized, normalized metrics (P3), 
    \item the amount of code needed to implement different models is greatly reduced (P3), 
    \item models are more reproducible (P15, P3),
    \item latest ML research can be easily incorporated into existing workflows (P3, P13),
    \item model training requires less human intervention as there are fewer errors (P2)
\end{itemize}
For hosted \automl solutions that generate extensive reports on the end-to-end process, \automl serves as a self-documenting, reproducible knowledge sharing platform (P10, P3). This is especially beneficial in industry, as P3 points out:
\begin{quote}
    ``Data scientists are expensive and very in demand, and people leave the job a lot and, and the algorithms change all the time. If I quit my job today \ldots I could be like, here's all the history.''
\end{quote}

\topic{Prevents suboptimal results due to idiosyncrasies of \experts}
A unique benefit of \automl 
applicable only to \experts is its ability to 
prevent suboptimal model performance resulting from idiosyncratic practices by \experts with extensive ML experience.
P5 relayed instances where the \automl tool found models that they never would have tried manually but outperformed the conventional choices they made. They therefore dubbed \automl the ``no assumptions approach'' to ML development.
P8 preferred the fact that the only factor driving the decisions made by the \automl tool is ``the statistical properties of the data'' and not their own familiarity of specific model types, thus removing ``bias'' from the process. 

\topic{Builds models more effectively and efficiently} 
The ability to \textit{build better models faster} is a major benefit of \automl tools reported by many participants, especially \intermediates.

Most participants reported that \automl led to significant improvements in \textit{efficiency}, the time for developing models, and a moderate increase in \textit{effectiveness}, the performance of the final model obtained. 
On a five-point Likert scale, participants on average rated improvement in efficiency ($\mu = 4.1$) higher than improvement in effectiveness  ($\mu = 3.88$). 
Some participants reported an order of magnitude reduction in development time (P10, P16).
Improvements in effectiveness experienced by the participants were often incremental; thus, participants did not deem more accurate models in isolation as a compelling reason for \automl adoption.

In addition to reduction in model development time, another dimension of efficiency is the ability to experiment with significantly more models in the same time it took for manual development. Even if the overall model performance does not change, 
exploring more models in itself provides downstream benefits. 
Many participants felt that they were much more productive because they were able to explore more models in the same amount of time. 
For P10, extensive experimentation also led to better insights into the features and models:
\begin{quote}
    ``The team that built out the manual process were only able to review  two or three different models. I was able to look at 50. 

    I was able to report on more insights, more understanding of the variable inputs than they were in the same amount of time, more understanding around why the model performed the way that it did.'' (P10)
\end{quote}

\topic{Enables rapid prototyping}
Traditionally, incorporating ML into an application from scratch is a lengthy process involving many stakeholders. The substantial overhead of adoption, on top of the uncertainty about whether ML will improve the application behavior, deters many from ML adoption.
\automl has significantly lowered the barrier to entry by enabling users to build quick prototypes to gauge the feasibility and potential impact of ML, without the cumbersome process of setting up infrastructure and codebase (P9, P11) and full integration (P12, P13), especially in industry settings.

The caveat is that for many industry users, \automl is used for rapid prototyping only but not full development, due to a lack of confidence in its performance (P4, P13)
\begin{quote}
    
    ``[I use Auto-ML] just to prototype and see how it works. 

    I don't use it within the system within my industry job, but I use it as a prototype system there just to see how easy this task is.'' (P4)
\end{quote}
\begin{quote}
    ``I would use \automl first to try things out, to understand. It would be a good starting point for a new application. I am convinced that it would generalize, but if I want a particular performance number, then I am less confident about how it would perform.'' (P13)
\end{quote}

\topic{Fosters Learning}

Users who were able to inspect the search history of the \automl tools reported that they learned about new modeling techniques (P8), implementation of specific ML algorithms (P9), model architecture (P11, P14), model performance on specific types of tasks (P10), and model resource consumption (P15). 
These learning opportunities emerged serendipitously, as the users were  validating the predictions and interpreting the models. 
However, for P2 who specifically sought to learn from their \automl tool $T$ (tool name anonymized to preserve participant privacy), the lack of transparency greatly hindered learning: 

\begin{quote}
    ``Getting the model out of $T$ proved to be extremely challenging. It was like 94\% accuracy 94\% precision. And when we tried that we didn't see anywhere close to that. 
    We tried to actually open up the model and see how it actually structured it, which was extremely challenging. It took a couple of hours, and then we learned that their structure was something that [they have written a paper on], so it was extremely hard to use. 

    \ldots It basically looks more like a black box.''
\end{quote}
Prior to their experience with $T$, P2 regarded $T$ highly on account of the cutting edge ML algorithms that $T$ claims to incorporate. However, due to their frustrations with the blackbox nature of the tool and the lack of offline reproducibility, P2 eventually abandoned \automl and reverted back to manual ML development.

%% file: 4_s4_deficiencies.tex

In this section, we present findings on the deficiencies of existing \automl tools that can potentially be addressed via systems innovations. 
We discuss the design implications of our findings in Section~\ref{sec:discussions}. 
There are other limitations of \automl tools that stem from the complex social and psychological implications 
of human-machine collaboration 
in ML workflows, 
and we discuss 
these limitations in Section~\ref{sec:humanRole}.

\topic{Lacks comprehensive end-to-end support}
Figure~\ref{fig:tools} shows that \textit{completeness}, the extent to which \automl covers the end-to-end ML workflow requirements, 
is the second lowest scoring rating. 
As evident in Figure~\ref{fig:use_cases}, 
\automl is currently used primarily 
for automating model training, requiring users to do the 
heavy lifting for both data pre-processing and 
post-processing using other tools. 
This reality directly contradicts 
some claims made by \automl tool developers. 
In the \automl platform category:
\begin{quote}
    ``[DataRobot] supports all of the steps needed to prepare, build, deploy, monitor, and maintain powerful AI applications at enterprise scale. [It] even automates model deployment, monitoring, and management.''
\end{quote}
\begin{quote}
    ``[H2O Driverless AI delivers] automatic feature engineering, model validation, model tuning, model selection and deployment, machine learning interpretability, bring your own recipe, time-series and automatic pipeline generation for model scoring.''
\end{quote}

Cloud and OSS solution developers are less aggressive in claiming end-to-end support, since OSS could rely on programmatic interoperability with other libraries, and cloud providers in theory could integrate with their other offerings for other stages of the ML workflow. 
While intended for flexibility, 
the interoperable design led to a \textit{fragmented data ecosystem},
causing users to \textit{``spend most of the time gluing everything together.''} (P14)
In practice, all participants who used cloud-hosted \automl reported using it in isolation, 
necessitating significant manual effort for data ingress and egress. 
P12, user of a cloud-based \automl tool $C$ lamented: 

\begin{quote}
    ``The biggest challenge is \ldots manipulating the data in a way that can be used with [$C$]. That is not something that we would have done if not for $C$. \ldots For each project you'll have to spend considerable amount of time structuring the data in a way that can be fed into $C$.'' (P12)
\end{quote}
This drawback places cloud solutions below OSS in participant-rated completeness, despite the fact that cloud solutions provide more built-in features for model deployment and interpretability. 

\textit{Limited data pre-processing.}
In terms of functionalities, a common complaint across all tools categories is inadequate support for data wrangling.
As P1 pointed out, data pre-processing support in \automl tools is primarily for feature engineering, which is deemed satisfactory by many \automl users, and does not cover data cleaning and 
wrangling needs. In fact, P1, P3, and P15 stated that their \automl tools did not support data pre-processing even though they acknowledged the feature engineering functionalities of their tools, since most of their time is spent on data wrangling and not feature engineering. P6's choice of \automl tool was determined 
entirely by the tool's ability to 
automate domain specific data wrangling.

Among the host of data wrangling tasks enumerated in Section~\ref{sec:useCaseDPR}, participants expressed the need for improved system support for the following tasks:
\begin{itemize}
    \item automated data discovery (P1)
    \item data cleaning for domain specific data (P6) and with more user control (P7) to avoid ``garbage in, garbage out''
    \item data transformation for domain specific data (P9, P10, P13) and for mitigating common data problems such as class imbalance (P16)
    \item large-scale data processing using distributed architecture (P14)
    \item dataset augmentation using state-of-the-art research (P11).
\end{itemize}

\textit{Limited Support for complex models and data types.}
Classification and regression 
are currently the only types of tasks supported by \automl tools. While some tools are beginning to support unstructured data such as text and images by harnessing recent development in deep learning, tabular data remains the focus for most \automl tools. The existing user-base for \automl is self selected 
to fit into the capabilities of current offerings. Even so, many participants expressed the desire for more broad-ranging support, such as unsupervised learning (P3, P6, P10, P15, P16) and domain-specific data models (P6: healthcare, P10: time series). 

\topic{Causes system failures due to compute intensive workloads}
A common complaint about \automl by participants 
who used OSS solutions is system performance, 
a major contributing factor to OSS being 
rated lower than the other categories of 
tools on \textit{ease of use} in Figure~\ref{fig:tools}.
P14 reported that \textit{``running out of 
main memory was the biggest technical challenge.''}
P8, whose models had \textit{18 million columns}, 
also reported that running out of memory, 
which crashed model training, was a frequent frustration. 
P5 had to run experiments on limited computation resources provided to her research lab, and she had to modify the model search space to reduce the total run time:
\begin{quote}
    ``[It was] time consuming for [\automl] for large dataset, and some pipelines are just too heavy and crash the process. For large feature set and sample set, some operators to further expand the data set were [slow and had to be] removed from the search space.'' (P5)
\end{quote}
P11, P14, and P15 also reported that they needed to define the model search space carefully to cope with the compute-intensive nature of \automl workloads, and P15 would even revert back to manual development for large models. The need to switch between development on laptops and on servers posed a major \textit{``annoyance''} for P11.

P5, P8, P11, P14, and P15 all reported spending 
only a fraction of their 
time on model development, 
thus their heavy \automl compute needs are intermittent.

\topic{Lacks customizability}
Customizability is the third 
lowest rated quality of \automl tools by the participants, 
as shown in Figure~\ref{fig:tools}. 
Interestingly, both too little customizability and too much customizability contributed to the low rating for customizability. 

Wanting more customizability is a sentiment shared by many \intermediates and \experts, especially users of cloud-hosted solutions. Participants wanted more custom control for computation resource allocation (P2, P4), data cleaning procedures (P6, P7), model search space (P4, P13), and model interpretability techniques (P5). 

On the other hand, too many customization options could lead to cognitive overload and hinder progress. P15 reported feeling overwhelmed by the number of hyperparameters that could be customized and needed to consult the documentation. 
P9, a novice, shared that they believed ``there's a lot of tools higher than a five [for customizability] but in a bad way.''
P3 described the phenomenon of gratuitous customizability:
\begin{quote}
    ``You don't necessarily know what some of the hyperparameters mean some of the time in extensive detail, but you do have the ability to control them all.''
\end{quote}

Most notably, P16 realized during the course of the interview that the additional customizability they wanted was in fact unnecessary for their use case:
\begin{quote}
    ``I've kind of been harping on how it's not as customizable \ldots But the tools that I've looked at lets you select the types of models to evaluate and change your features.
    They give you capability of managing the the information and shaping the underlying model \ldots It's handled quite well.''
\end{quote}
We will delve deeper into the issue of customizability and control in Section~\ref{sec:humanRole}.

\topic{Lacks Transparency and Interpretability}
The lowest rated quality of \automl tools is transparency for the cloud solutions due to their black-box nature and relative lack of opportunities for user agency in comparison to other tool categories. For increased transparency and usability,
a couple of participants expressed the desire for 
a simple progress bar that gives 
them insights into how long \automl would take (P2, P13.) However, different user populations desire different levels of transparency to ensure trust:  
\begin{quote}
    ``\automl is also an ML model. What that ML model is, how the ML model was trained, how the ML model learns from newer data---that piece is a black box. And so that makes it less trustworthy for people like me who are also ML 
    engineers who knows the success probability of machine learning models. For someone who's not in ML, it's like magic. You just click a button and it works. But for someone who knows ML \ldots I know one of the things that can go wrong [is] a ML model that is not trained well. 
    One of the problems \ldots with \automl is that they don't give you a lot of information about what is actually going on behind the scenes, and that makes it really hard for me to trust.'' (P12)
\end{quote}

Although widely reported by prior work on human-centered \automl \cite{Drozdal2020TrustAutoML,wang2019atmseer} that transparency mechanisms increase user trust in \automl systems, our results indicate that transparency mechanisms alone, 
such as visualization, does not suffice for the level of trust 
required in  high-stake industrial settings, 
wherein participants need to reason and justify for how and why the model design and selection decisions are made. In order to gain the level of interpretability and trust required for mission-critical projects, participants reported switching to complete manual development for increased user agency and control (P1, P4, P9, P11, P12, P15), as further discussed in Section~\ref{sec:humanRole}.
Conversely, lack of agency and tinkering can result 
in a lack of interpretability and the type of non-transparency caused by illiteracy~\cite{Burrell2016Opacity}. Humans develop understanding by doing, as illustrated by P3:
\begin{quote}
    ``You don't have a fundamental understanding of what's happening under the hood. And the other challenges with that \ldots are interpretability \ldots The onus is on me to actually build a competency in them \ldots It makes it basically impossible to go to a business person \ldots [to explain] how do you decide on this transaction. \ldots I actually have to go back \ldots and look at the 300 algorithms documentation, not the best one that I just deployed without really reading up all the details. There is a lot that you let go.''
\end{quote}

%% file: 4_s5RoleHuman.tex

Contrary to the moniker ``automated ML'', practitioners do not use \automl tools as push-button, one-shot solutions. Instead, they get into the trenches with \automl during the model development process---instructing, advising, and safeguarding \automl. Humans are valuable contributors, mentors, and supervisors to \automl, improving its efficiency, effectiveness, and safety. Their place ``in the loop'' cannot be replaced with complete automation. The consequences of removing humans out of the process are ineffectiveness, unpredictability, and potential danger. Below we discuss the crucial roles humans play alongside \automl.  

\subsubsection{Humans boost \automl's performance and efficiency} 
\automl, 
despite all its benefits, 
cannot be efficient and effective 
without humans-in-the-loop, because humans' contextual awareness, 
domain expertise, and ML experience
cannot be automated away. 
Humans are especially indispensable 
for non-standard uses cases and domains. 
Without human involvement, \automl cannot 
meet the stringent requirements of real-world applications. 

\topic{Human guidance constricts the search space of \automl}
The flip side of \automl's comprehensiveness 
(as a benefit in Section~\ref{sec:benefits}) 
is the enormous search space that requires an 
unwieldy amount of compute and time resources, 
when attempting to maximize model performance (accuracy, or other performance metrics). 
Due to \automl's exhaustive search process,
participants often only use \automl for light and basic models 
that can be trained very quickly 
with a data set that is not too big (P9, P11, P14, P15). 
\automl blindly searches through the entire space possible, resulting in intractable compute requirements, where ``you will be there for years and years searching for models. It's not possible.'' (P11). Paradoxically, present day \automl tools have no memory of and do not learn from the previous searches to narrow down the space in the next iteration, as P11 described:
    \begin{quote}
        ``But [\automl] doesn't learn how to learn, and it doesn't learn the environment in which it learns \ldots I could put in more samples if I want to iterate again \ldots I have to do a whole new \automl search, which is much more time consuming.'' 
    \end{quote}
    
Humans, in contrast, learn from their previous experiences, 
and can guide \automl using heuristics, thus narrowing the search space for \automl, 
so that it can return outputs that meet the quality standards 
within time constraints. 
The most prevalent strategy that participants applied is to define the inclusion and/or exclusion criteria for models and/or hyperparameters (P7, P8, P12.)  
For example, P8 reported having to curate a list of models 
to feed into \automl to constrain the search space. 
P14 also discussed the importance
of manually limiting the hyperparameter search space:
    \begin{quote}
    ``I think one problem with \automl is that it's very compute intensive. So you actually need to \textit{define your space in a good way.} For example, if you want to find the range of hyperparameters, you need to have some kind of notion of what you want to use as initial parameters. So basically you have a trade-off between time and model accuracy.'' (P14)
    \end{quote}
P14's use case requires them to hold the time constraint 
as a constant, wherein model performance and search space size are inversely related, because for large search spaces untamed by human guidance, \automl would `cut corners' by skipping some search areas, weakening its performance. P9 described their mental model of \automl's search process when they configured a time limit on \automl runs: 
    \begin{quote}
       ``It (\automl) wouldn't necessarily do hyperparameter tuning. I'm only realistically going to be able to do that, if I'm using simpler models. Otherwise, I have to sit down with someone who knows the models really well and get a good default set of hyperparameters.''
    \end{quote}

\topic{Humans compensate for \automl shortcomings, boosting its performance}
\automl's shortcomings become more evident 
in non-standard use cases and domains (P4, P6, P9, P11, P12). 
In such cases, humans use \automl 1) to establish an baseline for performance scores 
2) to learn from \automl's strategies~\ref{sec:benefits} 
and manually improve ML model performance. For example, P4 reported: 
    \begin{quote}
        ``When the task gets too complicated, \automl breaks down \ldots you won't get good state-of-the-art results. As soon as the task becomes something out of the normal, I switch to manual. \automl is just a good way to get the intuition behind what kind of performance you should expect in these cases. And then you try to beat that.'' 
    \end{quote}
\automl also often ``over-thinks'' and ``over-complicates'', resulting in diminished performance. Some participants respond by comparing manual and \automl models and objectively select the best performing one based on certain performance metrics.  
    \begin{quote}
        ``A couple of times, I ended up not using \automl, because it was giving me a very complicated pipeline for regression problem. So I limited it to elastic net only, and that works better than what \automl gave me. \ldots I select the best ones among manual and auto models.''(P5)
    \end{quote}
    
Many participants describe their working relationship with \automl as if they are collaborators~\cite{Dakuo2019Collaborative}, working in alternating cycles, iterating based on the feedback from each other (P10, P11, P13.) Humans leverage \automl's strengths and compensate for its shortcomings, engaging with \automl when it needs help. 
Together, they achieve higher performance (in speed and accuracy) 
than if they were each to work on their own. 
    
For example, P10 jump-starts their ML workflow with a quick \automl run to narrow down the variables/features, followed by manual feature engineering and refinement
via several iterations with \automl before finalizing the training sample 
for \automl to build models with. 
    \begin{quote}
        ``\textit{What types of pre-processing is \automl automating for you?}
        Oh, it's just sample selection. We can quickly run through modeling exercises and see which features or variables that we've tossed in are most important, and we're going to start chopping things out \ldots really quickly. And then go back to potentially doing some additional feature engineering ourselves manually or pulling more data in, but on a much limited scope \ldots And I'm going to continue to refine with multiple iterations of modeling, what ultimately I'm going to use as a training sample.'' (P10)
    \end{quote}
P10 compensates for \automl's shortcoming in feature engineering, while also supporting \automl with data collection---one of the tasks that are extremely to automate.  
    
\topic{Humans do what \automl cannot do at all, using contextual awareness and domain expertise} 
As reported in Section ~\ref{fig:use_cases}, many participants do not believe that tasks such as data pre-processing can ever be automated. The most common steps in the ML workflow where humans are indispensable are data collection (including data labeling), data cleaning, and feature engineering.(P1, P2, P3, P5, P6, P9, P10, P12, P14). 
Some ML tasks are extremely difficult to automate, such as unsupervised learning and semi-supervised learning (P7, P16.) 
Certain common data types also require substantial domain expertise, such as text (P9.)
Automation's rigidity and lack of contextual awareness are well-studied in HCI~\cite{Ackerman2000SocialTechnicalGap}. The same is true for \automl. 
As P6 puts it ``\automl is not smarter. It doesn't do what humans cannot do. It is just faster.''

\subsubsection{Humans increase ML safety and prevent misuse of \automl}  
\automl's ease-of-use also becomes its Achilles' heel. 
Participants reported \automl's excessive ease-to-use 
to be a downfall (P9, P15). 
P10 suggested that \automl is effective ``under proper governance and management.''
P9 responded to the question ``If an \automl and a manual model have the same performance, 
in which model would you have more confidence?'' with ``It depends on the person who used \automl.'' 
To ensure the safety of \automl decisions and prevent misuse, 
participants actively engage in the entire workflow as supervisors of \automl, implementing governance and management. The common strategies include the following:

\topic{Humans compare manual and \automl strategies as a safety check and to increase trust} 
Because of the shortcomings of \automl, participants expressed the need to personally validate \automl to establish confidence in \automl models, which often entails manually developing models to compare with \automl outputs (P5, P10, P11, P12, P16.)  
In addition, practitioners often engage with \automl 
with a prior expectation of performance. As P16 reported: 
    \begin{quote}
        ``There are certain times where running your own model, even if it's just as another perspective on the approach, and a \textbf{validation step} is still a good idea \ldots making sure that the direction of the predictions are in line with expectations [and] there's nothing abnormal happening.''(P16)
    \end{quote}
Prior work found that ML interpretability seeks 
to establish trust not only between humans and models, but also between humans~\cite{hong2020human}. 
Participants reported comparing \automl to 
manual ML for building trust between people. 
    \begin{quote}
        ``My goal is to bring to my collaborators interpretable models \ldots the biggest challenge [with \automl] is how to convince my collaborators to trust it. [The way I convince them is] we produce the standard approach models and show them that this one (\automl) is actually better in performance terms.'' (P5)
    \end{quote}
    
\topic{Humans correct for \automl idiosyncrasies} 
Ironically, as much as \automl 
prevents suboptimal performance 
due to practitioners' idiosyncrasies 
(as mentioned in Section~\ref{sec:benefits}), humans correct for 
\automl's idiosyncracies  
to safeguard \automl outputs 
and improve overall performance. 
P10 describes one such issue with data leakage:
    \begin{quote}
        ``The only way that [\automl ]detects target leakages is if you have like a 99.99\% correlation to the target label. So there's still things you need to know as a data scientists to use \automl effectively.''
    \end{quote}
P10 manually structured the data to ensure \automl has the appropriate data as input to safeguard the model's generalizability and safety.
    
P16 worked with heavily imbalanced data and had to manually correct the biases in the data to prevent suboptimal performance before feeding the dataset to \automl.    
    \begin{quote}
    ``I know \automl features allow you to do variable weighting. But \ldots I haven't found that to always work necessarily. ''(P16)
\end{quote}

\topic{Humans manually develop ML for understandability and reliability for mission-critical projects} 
Visibility into \automl work process does not suffice for the level of understanding, trust, and explainability participants need for high-stake projects, where humans are ultimately accountable for the reliable performance of ML models. Participants reported switching back to manually developing ML pipelines for mission-critical settings, because they need to reason and justify why the architecture and hyperparameters are chosen and how classification or prediction decisions are arrived at (P1, P4, P9, P11, P12, P15.) 
    
Prior work concludes that transparency mechanisms, such as visualization, increase user understanding and trust in \automl~\cite{Drozdal2020TrustAutoML}. We found that transparency alone does not suffice for trust and understanding between humans and the tool-built model. Humans need agency to establish the level of understanding to trust \automl. For example, P11, despite having visibility into \automl, was afforded little understanding. P11 rated transparency of \automl as being extremely transparent. 
    \begin{quote}
        ``\textit{To what extent do you have visibility into the inner workings of what the \automl is doing?}
        If you want, it is a five (extremely visible). You can log whatever you want. And you can see what it's doing and it saves all the run and you can look at them in tensor board, you can even explore the architectures, so extremely [visible].'' 
    \end{quote}
However, P11 distrusted in \automl and resorted to manual involvement. 
    \begin{quote}
        ``[With] \automl models, I can look at the architecture and I got no idea what it's doing. It is making connections all over the place. It is using non-conventional convolutional layers and you're like this is not predictable. Granted it's got good performance on one run \ldots You can't afford to have that kind of uncertainty \ldots You haven't had the \textbf{hands-on experience} to at least put your signature on the models and say put into production, so that's the issue.''
    \end{quote}
    
P11 reported that lack of certainties and predictability in how \automl selects model architecture undermined their trust in \automl and that human agency increased model reliability and assures confidence in the model. This sentiment is also supported by other participants. For example P7, P13 prefer manual development for use cases where \automl's strength in efficiency is irrelevant, because through manual ML development they can gain a deeper understanding and higher trust in the models.

%% file: 5_discussions.tex

\section{Discussion}
\label{sec:discussions}
Our findings show that \automl tools have been effective at making ML more accessible by creating high-level abstractions that streamline the process for training and deploying models. By effectively hiding away the complexities commonly associated with ML and acting as a source of knowledge, \automl tools make ML accessible to novices and increase productivity for experts. However, we argue that current efforts in \automl that strive for an eventual fully automated system through more tool capabilities and more accurate models fail to consider the ways the technology is actually used in practice and what users really need. In this work, we found that \textit{complete automation is neither a requirement nor a desired outcome for users of \automl}. To make \automl tools more effective, the goal should {\em not} be to completely remove the user from the process but instead to build human-compatible tools by working to achieve trust, understanding, and a sense of agency in the user. In this section, we discuss directions for the further development of \automl tools based on our findings.

\topic{Adapt to the proficiency level of the intended user} 
Especially because \automl tools mainly target citizen data scientists who often are intimated by ML due to perceived self-inefficacy, developers need to consider users' psychological readiness, designing from the position of a partner who has a sensitivity to the other's level of comfort (P12, P16.)  P16 expressed this as follows: 
\begin{quote}
    ``A lot of the \automl tools make an assumption about the technical competency of the user. That's to their detriment. I think if the goal is really to try and make it easier to use, there needs to be substantial effort put into the UX, and understanding of potential users that don't have\ldots the depth of knowledge\ldots [It is used] more as exploratory analytics or proof of concept.'' 
\end{quote}
Therefore \automl tools need to be designed to adapt to users with varying comfort level instead of taking a one-size-fits-all approach.

\topic{Lingua Franca for ML} 
Developers of \automl tools need to grapple with all the different roles that humans play in real-world data science work practices and how tools often need to take on a translator role among collaborators. P1 reported that even though many data science projects are initiated by business teams, the data scientists also often proactively propose innovative solutions to business teams based on their awareness of the challenges faced by business teams. ML engineers in real-world working environments do not simply passively react to requests from stakeholders. They also make contributions to the framing of the problem based on their ML expertise. Prior work highlights a translator role who sits between the data scientists and other stakeholders in collaborative data science projects~\cite{Dakuo2019Collaborative}. We found that many ML engineers perform the role of the translator, directly interfacing with stakeholders, interpreting business problems and framing them into objectives that can be evaluated by ML. P9 described this challenge of mismatched mental models among collaborators:
\begin{quote} 
    ``The way business thinks about accuracy is often really really different from how you would calculate any sort of traditional metrics\ldots You have to sync with a lot of people on exactly how they think accuracy works and how they want to report it.''
\end{quote}
\automl tools need to be aware of this mismatch and adapt their language accordingly to proactively act as a translator.

\topic{Holistic platform} 
An important lesson to be learned from the large difference in favorability between the different categories of \automl tools is that an \textit{end-to-end} solution that handles all stages of the ML workflow in a single environment is the optimal design choice for \automl solutions.
As mentioned in Section~\ref{sec:tools}, \automl platforms tend to have self-sufficient end-to-end ML workflow support due to limited options to integrate with external solutions, unlike OSS and Cloud Provider solutions. 
This has led to \automl platforms being not only the most complete solution but also the easiest to use, the most efficient (no data transfer), and the most interpretable (comprehensive data lineage). 
Since \automl is a highly complex system catered towards users with diverse backgrounds performing cognitively taxing tasks, interoperability with external solutions is an anti-pattern. 
This is not to say that there is no use for non-platform \automl solutions, but rather there needs to be a common substrate for the disparate \automl tools and libraries, akin to Weld for data analytics~\cite{palkar2017weld}.

\topic{Serverless Computing} 
As presented in Section~\ref{sec:deficiencies}, \automl workloads are compute intensive but bursty, and the optimal hardware is highly variable depending on the dataset and model characteristics. These conditions motivate the need for elastic, ephemerally provisioned computation resources with variable specs. 
For example, for P8 whose dataset contained 18 million columns, they could be temporarily allocated machines with ample RAM, instead of the fixed-architecture university cluster they were using, to avoid crashing due to out-of-memory errors and accelerate model search. Recent advancements in serverless computing~\cite{baldini2017serverless} can be harnessed to solve this problem.
However, the \automl setting poses new challenges in terms of the economics of trading compute cost for model accuracy.

\topic{Adaptive UI} 
Supporting users with diverse skills and expertise is an inherent challenge \automl solutions must embrace. 
Evidence from Section~\ref{sec:benefits} suggests that a blackbox interface for \automl can be beneficial to \novices and lower the maintenance overhead for all, but low customizability and transparency associated with blackbox interfaces hinder trust and agency based on evidence from Section~\ref{sec:deficiencies}.
\automl tool developers are forced to grapple with competing design objectives, and a natural solution to this conundrum is to provide multi-modal interfaces covering a spectrum of interaction levels. 

The \automl tooling landscape has progressed towards the low code/no code direction, with some solutions allowing experienced users to assume more control via a secondary programmatic interface. However, these tools do not offer true multi-modal interfaces but merely the option to export the raw model training code for further manual exploration, creating a clunky user interface ``geared towards ML engineers who can learn to deal with broken UIs'' (P12). 
Furthermore, the success of a multi-modal interface is contingent upon a user's ability to self select the most appropriate modality, which is not easily achievable as illustrated by P16's experience with customizability. 
A possible solution to this UX challenge is instead of having multiple distinct modalities, the \automl tool can adaptively reveal or hide customization capabilities piecemeal, based on some approximation of user skill level and intent.

\topic{Interactive Exploration} 
Our findings in Section~\ref{sec:humanRole} show that the human ``in-the-loop'' can be an indispensable resource to complement \automl and make it more efficient and effective. In tools that support search space customization, human supervision is provided in a one-shot fashion. If the user wanted to iteratively refine the search space, they would have to manually kick off multiple rounds of the one-shot process, keeping track of intermediate results manually between iterations. 
Iterating with \automl can be better supported by an interactive exploration interface designed specifically with iteration in mind. Such an interface needs to display summaries of all previous iterations and make it easy for the user to specify new search sub-spaces. It also needs to reconcile the high latencies of \automl workloads and interactivity.

\topic{Balance between Human Control and Automation}
In Section~\ref{sec:results}, we presented several deficiencies of the tools that can be improved with some of the suggestion above, as well as roles that are important for the human to assume when using \automl. The decisions for what functionalities belong in either group are predicated upon, first and foremost, how to establish trust and agency in the human users, and the current state of ML, systems, and HCI research. Depriving users of trust and agency prevents \automl from making an impact in real-world use cases, while attempting to automate certain features prematurely, without fundamental shifts in the underlying technology, requires strenuous efforts with little payoff in the user experience. 

Certain tasks cannot be addressed with human control or automation alone but rather require a delicate balance between the two. Take data pre-processing for example. Evidence in Section~\ref{sec:humanRole} suggests that many users deem complex feature engineering simply out-of-reach for existing Auto-ML systems, due to their inability to capture domain knowledge. However, many also feel that existing tools lack basic support for mechanical tasks such as distributed processing and canonical data transformations (Section~\ref{sec:deficiencies}). Thus, the future of data preprocessing is a combination of adding support for mechanical tasks in the near term, and providing intuitive ways to specify high-level domain knowledge to integrate human intuition with \automl long term.

Another example is the collaboration of humans and the machine for efficient hyperparameter tuning. While expert users believe that they can improve the efficiency of \automl through intuition and experience, this belief is juxtaposed with their recognition that automated search can also correct for their biases and idiosyncrasies (Section~\ref{sec:benefits}). To strike a balance, we envision an interactive dialog between the human and the machine to iteratively discover and fill in each other’s blindspots. Human guidance will be treated as strong priors on certain model subspaces but does not preclude the exploration of other subspaces, allowing the machine to nudge the human towards promising but underexplored subspaces.

%% file: main.bbl

\begin{thebibliography}{48}


\ifx \showCODEN    \undefined \def \showCODEN     #1{\unskip}     \fi
\ifx \showDOI      \undefined \def \showDOI       #1{#1}\fi
\ifx \showISBNx    \undefined \def \showISBNx     #1{\unskip}     \fi
\ifx \showISBNxiii \undefined \def \showISBNxiii  #1{\unskip}     \fi
\ifx \showISSN     \undefined \def \showISSN      #1{\unskip}     \fi
\ifx \showLCCN     \undefined \def \showLCCN      #1{\unskip}     \fi
\ifx \shownote     \undefined \def \shownote      #1{#1}          \fi
\ifx \showarticletitle \undefined \def \showarticletitle #1{#1}   \fi
\ifx \showURL      \undefined \def \showURL       {\relax}        \fi
\providecommand\bibfield[2]{#2}
\providecommand\bibinfo[2]{#2}
\providecommand\natexlab[1]{#1}
\providecommand\showeprint[2][]{arXiv:#2}

\bibitem[\protect\citeauthoryear{Ackerman}{Ackerman}{2000}]%
        {Ackerman2000SocialTechnicalGap}
\bibfield{author}{\bibinfo{person}{Mark~S. Ackerman}.}
  \bibinfo{year}{2000}\natexlab{}.
\newblock \showarticletitle{The Intellectual Challenge of CSCW: The Gap between
  Social Requirements and Technical Feasibility}.
\newblock \bibinfo{journal}{\emph{Hum.-Comput. Interact.}}
  \bibinfo{volume}{15}, \bibinfo{number}{2} (\bibinfo{date}{Sept.}
  \bibinfo{year}{2000}), \bibinfo{pages}{179–203}.
\newblock
\showISSN{0737-0024}
\urldef\tempurl%
\url{https://doi.org/10.1207/S15327051HCI1523_5}
\showDOI{\tempurl}


\bibitem[\protect\citeauthoryear{AI}{AI}{2017}]%
        {mlfacets}
\bibfield{author}{\bibinfo{person}{Google AI}.}
  \bibinfo{year}{2017}\natexlab{}.
\newblock \bibinfo{title}{Facets: An Open Source Visualization Tool for Machine
  Learning Training Data}.
\newblock
\newblock
\urldef\tempurl%
\url{https://ai.googleblog.com/2017/07/facets-open-source-visualization-tool.html}
\showURL{%
\tempurl}


\bibitem[\protect\citeauthoryear{AmazonSagemakerAutopilot}{AmazonSagemakerAutopilot}{2020}]%
        {AmazonSagemakerAutopilot}
\bibfield{author}{\bibinfo{person}{AmazonSagemakerAutopilot}.}
  \bibinfo{year}{2020}\natexlab{}.
\newblock \bibinfo{title}{Automate model development with Amazon SageMaker
  Autopilot}.
\newblock \bibinfo{howpublished}{Website}.
\newblock
\urldef\tempurl%
\url{https://docs.aws.amazon.com/sagemaker/latest/dg/autopilot-automate-model-development.html}
\showURL{%
Retrieved July 26, 2020 from \tempurl}


\bibitem[\protect\citeauthoryear{Amershi, Begel, Bird, DeLine, Gall, Kamar,
  Nagappan, Nushi, and Zimmermann}{Amershi et~al\mbox{.}}{2019}]%
        {Amershi2019Software}
\bibfield{author}{\bibinfo{person}{Saleema Amershi}, \bibinfo{person}{Andrew
  Begel}, \bibinfo{person}{Christian Bird}, \bibinfo{person}{Robert DeLine},
  \bibinfo{person}{Harald Gall}, \bibinfo{person}{Ece Kamar},
  \bibinfo{person}{Nachiappan Nagappan}, \bibinfo{person}{Besmira Nushi}, {and}
  \bibinfo{person}{Thomas Zimmermann}.} \bibinfo{year}{2019}\natexlab{}.
\newblock \showarticletitle{Software Engineering for Machine Learning: A Case
  Study}. In \bibinfo{booktitle}{\emph{Proceedings of the 41st International
  Conference on Software Engineering: Software Engineering in Practice}}
  \emph{(\bibinfo{series}{ICSE-SEIP ’19})}. \bibinfo{publisher}{IEEE Press},
  \bibinfo{address}{Montreal, Quebec, Canada}, \bibinfo{pages}{291–300}.
\newblock
\urldef\tempurl%
\url{https://doi.org/10.1109/ICSE-SEIP.2019.00042}
\showDOI{\tempurl}


\bibitem[\protect\citeauthoryear{Amershi, Cakmak, Knox, and Kulesza}{Amershi
  et~al\mbox{.}}{2014}]%
        {Amershi2014}
\bibfield{author}{\bibinfo{person}{Saleema Amershi}, \bibinfo{person}{Maya
  Cakmak}, \bibinfo{person}{William~Bradley Knox}, {and} \bibinfo{person}{Todd
  Kulesza}.} \bibinfo{year}{2014}\natexlab{}.
\newblock \showarticletitle{{Power to the People: The Role of Humans in
  Interactive Machine Learning}}.
\newblock \bibinfo{journal}{\emph{AI Magazine}} \bibinfo{volume}{35},
  \bibinfo{number}{4} (\bibinfo{year}{2014}), \bibinfo{pages}{105--120}.
\newblock
\showISSN{0738-4602}
\urldef\tempurl%
\url{https://doi.org/10.1609/aimag.v35i4.2513}
\showDOI{\tempurl}


\bibitem[\protect\citeauthoryear{Amershi, Chickering, Drucker, Lee, Simard, and
  Suh}{Amershi et~al\mbox{.}}{2015}]%
        {amershi2015modeltracker}
\bibfield{author}{\bibinfo{person}{Saleema Amershi}, \bibinfo{person}{Max
  Chickering}, \bibinfo{person}{Steven~M. Drucker}, \bibinfo{person}{Bongshin
  Lee}, \bibinfo{person}{Patrice~Y. Simard}, {and} \bibinfo{person}{Jina Suh}.}
  \bibinfo{year}{2015}\natexlab{}.
\newblock \showarticletitle{{ModelTracker: Redesigning Performance Analysis
  Tools for Machine Learning}}. In \bibinfo{booktitle}{\emph{{Proceedings of
  the 33rd ACM SIGCHI Conference on Human Factors in Computing Systems}}}.
  \bibinfo{publisher}{{ACM}}, \bibinfo{address}{Seoul, Korea},
  \bibinfo{pages}{337--346}.
\newblock
\showISBNx{978-1-4503-3145-6}
\urldef\tempurl%
\url{https://doi.org/10.1145/2702123.2702509}
\showDOI{\tempurl}


\bibitem[\protect\citeauthoryear{Bach, Ehrenberg, Fries, Wu, and R{\'e}}{Bach
  et~al\mbox{.}}{2017}]%
        {bach2017snorkel}
\bibfield{author}{\bibinfo{person}{Alexander Ratner Stephen~H Bach},
  \bibinfo{person}{Henry Ehrenberg}, \bibinfo{person}{Jason Fries},
  \bibinfo{person}{Sen Wu}, {and} \bibinfo{person}{Christopher R{\'e}}.}
  \bibinfo{year}{2017}\natexlab{}.
\newblock \showarticletitle{Snorkel: Rapid Training Data Creation with Weak
  Supervision}.
\newblock \bibinfo{journal}{\emph{Proceedings of the VLDB Endowment}}
  \bibinfo{volume}{11}, \bibinfo{number}{3} (\bibinfo{year}{2017}),
  \bibinfo{pages}{269--282}.
\newblock


\bibitem[\protect\citeauthoryear{Baldini, Castro, Chang, Cheng, Fink, Ishakian,
  Mitchell, Muthusamy, Rabbah, Slominski, et~al\mbox{.}}{Baldini
  et~al\mbox{.}}{2017}]%
        {baldini2017serverless}
\bibfield{author}{\bibinfo{person}{Ioana Baldini}, \bibinfo{person}{Paul
  Castro}, \bibinfo{person}{Kerry Chang}, \bibinfo{person}{Perry Cheng},
  \bibinfo{person}{Stephen Fink}, \bibinfo{person}{Vatche Ishakian},
  \bibinfo{person}{Nick Mitchell}, \bibinfo{person}{Vinod Muthusamy},
  \bibinfo{person}{Rodric Rabbah}, \bibinfo{person}{Aleksander Slominski},
  {et~al\mbox{.}}} \bibinfo{year}{2017}\natexlab{}.
\newblock \showarticletitle{Serverless computing: Current trends and open
  problems}.
\newblock In \bibinfo{booktitle}{\emph{Research Advances in Cloud Computing}}.
  \bibinfo{publisher}{Springer}, \bibinfo{address}{none},
  \bibinfo{pages}{1--20}.
\newblock


\bibitem[\protect\citeauthoryear{Bringer, Israeli, Shoham, Ratner, and
  R\'{e}}{Bringer et~al\mbox{.}}{2019}]%
        {bringer2019osprey}
\bibfield{author}{\bibinfo{person}{Eran Bringer}, \bibinfo{person}{Abraham
  Israeli}, \bibinfo{person}{Yoav Shoham}, \bibinfo{person}{Alex Ratner}, {and}
  \bibinfo{person}{Christopher R\'{e}}.} \bibinfo{year}{2019}\natexlab{}.
\newblock \showarticletitle{Osprey: Weak Supervision of Imbalanced Extraction
  Problems without Code}. In \bibinfo{booktitle}{\emph{Proceedings of the 3rd
  International Workshop on Data Management for End-to-End Machine Learning}}
  \emph{(\bibinfo{series}{DEEM'19})}. \bibinfo{publisher}{Association for
  Computing Machinery}, \bibinfo{address}{New York, NY, USA}, Article
  \bibinfo{articleno}{4}, \bibinfo{numpages}{11}~pages.
\newblock
\showISBNx{9781450367974}
\urldef\tempurl%
\url{https://doi.org/10.1145/3329486.3329492}
\showDOI{\tempurl}


\bibitem[\protect\citeauthoryear{Burrell}{Burrell}{2016}]%
        {Burrell2016Opacity}
\bibfield{author}{\bibinfo{person}{Jenna Burrell}.}
  \bibinfo{year}{2016}\natexlab{}.
\newblock \showarticletitle{How the machine ‘thinks’: Understanding opacity
  in machine learning algorithms}.
\newblock \bibinfo{journal}{\emph{Big Data \& Society}} \bibinfo{volume}{3},
  \bibinfo{number}{1} (\bibinfo{year}{2016}),
  \bibinfo{pages}{2053951715622512}.
\newblock
\urldef\tempurl%
\url{https://doi.org/10.1177/2053951715622512}
\showDOI{\tempurl}
\showeprint{https://doi.org/10.1177/2053951715622512}


\bibitem[\protect\citeauthoryear{Cortes, Gonzalvo, Kuznetsov, Mohri, and
  Yang}{Cortes et~al\mbox{.}}{2017}]%
        {cortes2017adanet}
\bibfield{author}{\bibinfo{person}{Corinna Cortes}, \bibinfo{person}{Xavi
  Gonzalvo}, \bibinfo{person}{Vitaly Kuznetsov}, \bibinfo{person}{Mehryar
  Mohri}, {and} \bibinfo{person}{Scott Yang}.} \bibinfo{year}{2017}\natexlab{}.
\newblock \bibinfo{title}{AdaNet: Adaptive Structural Learning of Artificial
  Neural Networks}.
\newblock
\newblock
\showeprint[arxiv]{cs.LG/1607.01097}


\bibitem[\protect\citeauthoryear{Datarobot}{Datarobot}{2020}]%
        {Datarobot}
\bibfield{author}{\bibinfo{person}{Datarobot}.}
  \bibinfo{year}{2020}\natexlab{}.
\newblock \bibinfo{title}{DataRobot Automated Machine Learning}.
\newblock \bibinfo{howpublished}{Website}.
\newblock
\urldef\tempurl%
\url{https://www.datarobot.com/platform/automated-machine-learning/}
\showURL{%
Retrieved July 18, 2020 from \tempurl}


\bibitem[\protect\citeauthoryear{dedoose}{dedoose}{2020}]%
        {dedoose}
\bibfield{author}{\bibinfo{person}{dedoose}.} \bibinfo{year}{2020}\natexlab{}.
\newblock \bibinfo{title}{Dedoose}.
\newblock \bibinfo{howpublished}{Website}.
\newblock
\urldef\tempurl%
\url{https://www.dedoose.com}
\showURL{%
Retrieved September 15, 2020 from \tempurl}


\bibitem[\protect\citeauthoryear{Drozdal, Weisz, Wang, Dass, Yao, Zhao, Muller,
  Ju, and Su}{Drozdal et~al\mbox{.}}{2020}]%
        {Drozdal2020TrustAutoML}
\bibfield{author}{\bibinfo{person}{Jaimie Drozdal}, \bibinfo{person}{Justin
  Weisz}, \bibinfo{person}{Dakuo Wang}, \bibinfo{person}{Gaurav Dass},
  \bibinfo{person}{Bingsheng Yao}, \bibinfo{person}{Changruo Zhao},
  \bibinfo{person}{Michael Muller}, \bibinfo{person}{Lin Ju}, {and}
  \bibinfo{person}{Hui Su}.} \bibinfo{year}{2020}\natexlab{}.
\newblock \showarticletitle{Trust in AutoML: Exploring Information Needs for
  Establishing Trust in Automated Machine Learning Systems}. In
  \bibinfo{booktitle}{\emph{Proceedings of the 25th International Conference on
  Intelligent User Interfaces}} \emph{(\bibinfo{series}{IUI '20})}.
  \bibinfo{publisher}{Association for Computing Machinery},
  \bibinfo{address}{New York, NY, USA}, \bibinfo{pages}{297–307}.
\newblock
\showISBNx{9781450371186}
\urldef\tempurl%
\url{https://doi.org/10.1145/3377325.3377501}
\showDOI{\tempurl}


\bibitem[\protect\citeauthoryear{Duranton, Erlebach, Brégé, Danziger,
  Gallego, and Pauly}{Duranton et~al\mbox{.}}{2020}]%
        {GenderRatio}
\bibfield{author}{\bibinfo{person}{Sylvain Duranton}, \bibinfo{person}{Jörg
  Erlebach}, \bibinfo{person}{Camille Brégé}, \bibinfo{person}{Jane
  Danziger}, \bibinfo{person}{Andrea Gallego}, {and} \bibinfo{person}{Marc
  Pauly}.} \bibinfo{year}{2020}\natexlab{}.
\newblock \bibinfo{title}{What’s Keeping Women Out of Data Science?}
\newblock
  \bibinfo{howpublished}{\url{https://www.bcg.com/en-us/publications/2020/what-keeps-women-out-data-science}}.
\newblock


\bibitem[\protect\citeauthoryear{Fails and Olsen}{Fails and Olsen}{2003}]%
        {Fails2003}
\bibfield{author}{\bibinfo{person}{Jerry~Alan Fails} {and}
  \bibinfo{person}{Dan~R. Olsen}.} \bibinfo{year}{2003}\natexlab{}.
\newblock \showarticletitle{Interactive Machine Learning}. In
  \bibinfo{booktitle}{\emph{Proceedings of the 8th International Conference on
  Intelligent User Interfaces}} \emph{(\bibinfo{series}{IUI '03})}.
  \bibinfo{publisher}{Association for Computing Machinery},
  \bibinfo{address}{New York, NY, USA}, \bibinfo{pages}{39–45}.
\newblock
\showISBNx{1581135866}
\urldef\tempurl%
\url{https://doi.org/10.1145/604045.604056}
\showDOI{\tempurl}


\bibitem[\protect\citeauthoryear{Feurer, Klein, Eggensperger, Springenberg,
  Blum, and Hutter}{Feurer et~al\mbox{.}}{2015}]%
        {NIPS2015_5872}
\bibfield{author}{\bibinfo{person}{Matthias Feurer}, \bibinfo{person}{Aaron
  Klein}, \bibinfo{person}{Katharina Eggensperger},
  \bibinfo{person}{Jost~Tobias Springenberg}, \bibinfo{person}{Manuel Blum},
  {and} \bibinfo{person}{Frank Hutter}.} \bibinfo{year}{2015}\natexlab{}.
\newblock \showarticletitle{Efficient and Robust Automated Machine Learning}.
  In \bibinfo{booktitle}{\emph{Proceedings of the 28th International Conference
  on Neural Information Processing Systems - Volume 2}}
  \emph{(\bibinfo{series}{NIPS'15})}. \bibinfo{publisher}{MIT Press},
  \bibinfo{address}{Cambridge, MA, USA}, \bibinfo{pages}{2755–2763}.
\newblock


\bibitem[\protect\citeauthoryear{Fiebrink, Cook, and Trueman}{Fiebrink
  et~al\mbox{.}}{2011}]%
        {Fiebrink2011}
\bibfield{author}{\bibinfo{person}{Rebecca Fiebrink}, \bibinfo{person}{Perry~R
  Cook}, {and} \bibinfo{person}{Daniel Trueman}.}
  \bibinfo{year}{2011}\natexlab{}.
\newblock \showarticletitle{{Human Model Evaluation in Interactive Supervised
  Learning}}.
\newblock \bibinfo{journal}{\emph{CHI 2011}} \bibinfo{volume}{4},
  \bibinfo{number}{2} (\bibinfo{year}{2011}), \bibinfo{pages}{147--156}.
\newblock
\showISBNx{9781450302289}
\showISSN{23138734}
\urldef\tempurl%
\url{https://doi.org/10.14529/jsfi170202}
\showDOI{\tempurl}


\bibitem[\protect\citeauthoryear{Gil, Honaker, Gupta, Ma, D'Orazio, Garijo,
  Gadewar, Yang, and Jahanshad}{Gil et~al\mbox{.}}{2019}]%
        {gil2019towards}
\bibfield{author}{\bibinfo{person}{Yolanda Gil}, \bibinfo{person}{James
  Honaker}, \bibinfo{person}{Shikhar Gupta}, \bibinfo{person}{Yibo Ma},
  \bibinfo{person}{Vito D'Orazio}, \bibinfo{person}{Daniel Garijo},
  \bibinfo{person}{Shruti Gadewar}, \bibinfo{person}{Qifan Yang}, {and}
  \bibinfo{person}{Neda Jahanshad}.} \bibinfo{year}{2019}\natexlab{}.
\newblock \showarticletitle{Towards Human-Guided Machine Learning}. In
  \bibinfo{booktitle}{\emph{Proceedings of the 24th International Conference on
  Intelligent User Interfaces}} \emph{(\bibinfo{series}{IUI '19})}.
  \bibinfo{publisher}{Association for Computing Machinery},
  \bibinfo{address}{New York, NY, USA}, \bibinfo{pages}{614–624}.
\newblock
\showISBNx{9781450362726}
\urldef\tempurl%
\url{https://doi.org/10.1145/3301275.3302324}
\showDOI{\tempurl}


\bibitem[\protect\citeauthoryear{GoogleCloudAutoML}{GoogleCloudAutoML}{2020}]%
        {GoogleCloudAutoML}
\bibfield{author}{\bibinfo{person}{GoogleCloudAutoML}.}
  \bibinfo{year}{2020}\natexlab{}.
\newblock \bibinfo{title}{Google Cloud AutoML}.
\newblock \bibinfo{howpublished}{Website}.
\newblock
\urldef\tempurl%
\url{https://cloud.google.com/automl}
\showURL{%
Retrieved July 18, 2020 from \tempurl}


\bibitem[\protect\citeauthoryear{H2O.ai}{H2O.ai}{2020}]%
        {H2O.ai}
\bibfield{author}{\bibinfo{person}{H2O.ai}.} \bibinfo{year}{2020}\natexlab{}.
\newblock \bibinfo{title}{H2o.ai Automated Machine Learning}.
\newblock \bibinfo{howpublished}{Website}.
\newblock
\urldef\tempurl%
\url{https://docs.h2o.ai/h2o/latest-stable/h2o-docs/automl.html#automl-interface}
\showURL{%
Retrieved July 18, 2020 from \tempurl}


\bibitem[\protect\citeauthoryear{Hohman, Wongsuphasawat, Kery, and
  Patel}{Hohman et~al\mbox{.}}{2020}]%
        {hohman2020understanding}
\bibfield{author}{\bibinfo{person}{Fred Hohman}, \bibinfo{person}{Kanit
  Wongsuphasawat}, \bibinfo{person}{Mary~Beth Kery}, {and}
  \bibinfo{person}{Kayur Patel}.} \bibinfo{year}{2020}\natexlab{}.
\newblock \showarticletitle{Understanding and Visualizing Data Iteration in
  Machine Learning}. In \bibinfo{booktitle}{\emph{Proceedings of the 2020 CHI
  Conference on Human Factors in Computing Systems}}
  \emph{(\bibinfo{series}{CHI '20})}. \bibinfo{publisher}{Association for
  Computing Machinery}, \bibinfo{address}{New York, NY, USA},
  \bibinfo{pages}{1–13}.
\newblock
\showISBNx{9781450367080}
\urldef\tempurl%
\url{https://doi.org/10.1145/3313831.3376177}
\showDOI{\tempurl}


\bibitem[\protect\citeauthoryear{Hong, Hullman, and Bertini}{Hong
  et~al\mbox{.}}{2020}]%
        {hong2020human}
\bibfield{author}{\bibinfo{person}{Sungsoo~Ray Hong}, \bibinfo{person}{Jessica
  Hullman}, {and} \bibinfo{person}{Enrico Bertini}.}
  \bibinfo{year}{2020}\natexlab{}.
\newblock \showarticletitle{Human Factors in Model Interpretability: Industry
  Practices, Challenges, and Needs}.
\newblock \bibinfo{journal}{\emph{Proceedings of the ACM on Human-Computer
  Interaction}} \bibinfo{volume}{4}, \bibinfo{number}{CSCW1}
  (\bibinfo{year}{2020}), \bibinfo{pages}{1--26}.
\newblock


\bibitem[\protect\citeauthoryear{Jin, Song, and Hu}{Jin et~al\mbox{.}}{2019}]%
        {jin2019auto}
\bibfield{author}{\bibinfo{person}{Haifeng Jin}, \bibinfo{person}{Qingquan
  Song}, {and} \bibinfo{person}{Xia Hu}.} \bibinfo{year}{2019}\natexlab{}.
\newblock \showarticletitle{Auto-Keras: An Efficient Neural Architecture Search
  System}. In \bibinfo{booktitle}{\emph{Proceedings of the 25th ACM SIGKDD
  International Conference on Knowledge Discovery \& Data Mining}}
  \emph{(\bibinfo{series}{KDD '19})}. \bibinfo{publisher}{Association for
  Computing Machinery}, \bibinfo{address}{New York, NY, USA},
  \bibinfo{pages}{1946–1956}.
\newblock
\showISBNx{9781450362016}
\urldef\tempurl%
\url{https://doi.org/10.1145/3292500.3330648}
\showDOI{\tempurl}


\bibitem[\protect\citeauthoryear{{Kandel}, {Paepcke}, {Hellerstein}, and
  {Heer}}{{Kandel} et~al\mbox{.}}{2012}]%
        {Kandel}
\bibfield{author}{\bibinfo{person}{S. {Kandel}}, \bibinfo{person}{A.
  {Paepcke}}, \bibinfo{person}{J.~M. {Hellerstein}}, {and} \bibinfo{person}{J.
  {Heer}}.} \bibinfo{year}{2012}\natexlab{}.
\newblock \showarticletitle{Enterprise Data Analysis and Visualization: An
  Interview Study}.
\newblock \bibinfo{journal}{\emph{IEEE Transactions on Visualization and
  Computer Graphics}} \bibinfo{volume}{18}, \bibinfo{number}{12}
  (\bibinfo{date}{Dec} \bibinfo{year}{2012}), \bibinfo{pages}{2917--2926}.
\newblock
\showISSN{2160-9306}
\urldef\tempurl%
\url{https://doi.org/10.1109/TVCG.2012.219}
\showDOI{\tempurl}


\bibitem[\protect\citeauthoryear{Le, Fu, and Moore}{Le et~al\mbox{.}}{2020}]%
        {TPOT:Olson}
\bibfield{author}{\bibinfo{person}{Trang~T Le}, \bibinfo{person}{Weixuan Fu},
  {and} \bibinfo{person}{Jason~H Moore}.} \bibinfo{year}{2020}\natexlab{}.
\newblock \showarticletitle{Scaling tree-based automated machine learning to
  biomedical big data with a feature set selector}.
\newblock \bibinfo{journal}{\emph{Bioinformatics}} \bibinfo{volume}{36},
  \bibinfo{number}{1} (\bibinfo{year}{2020}), \bibinfo{pages}{250--256}.
\newblock


\bibitem[\protect\citeauthoryear{Lee, Xin, Lee, and Parameswaran}{Lee
  et~al\mbox{.}}{2020}]%
        {Lee2020}
\bibfield{author}{\bibinfo{person}{Angela Lee}, \bibinfo{person}{Doris Xin},
  \bibinfo{person}{Doris Lee}, {and} \bibinfo{person}{Aditya Parameswaran}.}
  \bibinfo{year}{2020}\natexlab{}.
\newblock \bibinfo{title}{Demystifying a Dark Art: Understanding Real-World
  Machine Learning Model Development}.
\newblock
\newblock
\showeprint[arxiv]{cs.LG/2005.01520}


\bibitem[\protect\citeauthoryear{Lee, Macke, Xin, Lee, Huang, and
  Parameswaran}{Lee et~al\mbox{.}}{2019}]%
        {Lee2019AHP}
\bibfield{author}{\bibinfo{person}{Doris Jung~Lin Lee},
  \bibinfo{person}{Stephen Macke}, \bibinfo{person}{Doris Xin},
  \bibinfo{person}{Angela Lee}, \bibinfo{person}{Silu Huang}, {and}
  \bibinfo{person}{Aditya~G. Parameswaran}.} \bibinfo{year}{2019}\natexlab{}.
\newblock \showarticletitle{A Human-in-the-loop Perspective on AutoML:
  Milestones and the Road Ahead}.
\newblock \bibinfo{journal}{\emph{IEEE Data Eng. Bull.}}  \bibinfo{volume}{42}
  (\bibinfo{year}{2019}), \bibinfo{pages}{59--70}.
\newblock


\bibitem[\protect\citeauthoryear{McDonald, Schoenebeck, and Forte}{McDonald
  et~al\mbox{.}}{2019}]%
        {McDonald2019InterraterReliability}
\bibfield{author}{\bibinfo{person}{Nora McDonald}, \bibinfo{person}{Sarita
  Schoenebeck}, {and} \bibinfo{person}{Andrea Forte}.}
  \bibinfo{year}{2019}\natexlab{}.
\newblock \showarticletitle{Reliability and Inter-rater Reliability in
  Qualitative Research: Norms and Guidelines for CSCW and HCI Practice}.
\newblock \bibinfo{journal}{\emph{Proceedings of the ACM on Human-Computer
  Interaction}}  \bibinfo{volume}{3} (\bibinfo{date}{11} \bibinfo{year}{2019}),
  \bibinfo{pages}{1--23}.
\newblock
\urldef\tempurl%
\url{https://doi.org/10.1145/3359174}
\showDOI{\tempurl}


\bibitem[\protect\citeauthoryear{McKinney et~al\mbox{.}}{McKinney
  et~al\mbox{.}}{2010}]%
        {mckinney-proc-scipy-2010}
\bibfield{author}{\bibinfo{person}{Wes McKinney} {et~al\mbox{.}}}
  \bibinfo{year}{2010}\natexlab{}.
\newblock \showarticletitle{Data structures for statistical computing in
  python}. In \bibinfo{booktitle}{\emph{Proceedings of the 9th Python in
  Science Conference}}, Vol.~\bibinfo{volume}{445}. \bibinfo{publisher}{SciPy
  2010}, \bibinfo{address}{Austin, TX}, \bibinfo{pages}{51--56}.
\newblock


\bibitem[\protect\citeauthoryear{MicrosoftAzureAutomatedML}{MicrosoftAzureAutomatedML}{2020}]%
        {MicrosoftAzureAutomatedML}
\bibfield{author}{\bibinfo{person}{MicrosoftAzureAutomatedML}.}
  \bibinfo{year}{2020}\natexlab{}.
\newblock \bibinfo{title}{Microsoft Azure Automated Machine Learning}.
\newblock \bibinfo{howpublished}{Website}.
\newblock
\urldef\tempurl%
\url{https://azure.microsoft.com/en-us/services/machine-learning/automatedml/}
\showURL{%
Retrieved September 15, 2020 from \tempurl}


\bibitem[\protect\citeauthoryear{Molino, Dudin, and Miryala}{Molino
  et~al\mbox{.}}{2019}]%
        {molino2019ludwig}
\bibfield{author}{\bibinfo{person}{Piero Molino}, \bibinfo{person}{Yaroslav
  Dudin}, {and} \bibinfo{person}{Sai~Sumanth Miryala}.}
  \bibinfo{year}{2019}\natexlab{}.
\newblock \bibinfo{title}{Ludwig: a type-based declarative deep learning
  toolbox}.
\newblock
\newblock
\showeprint[arxiv]{cs.LG/1909.07930}


\bibitem[\protect\citeauthoryear{Ono, Castelo, Lopez, Bertini, Freire, and
  Silva}{Ono et~al\mbox{.}}{2020}]%
        {ono2020pipelineprofiler}
\bibfield{author}{\bibinfo{person}{Jorge~Piazentin Ono}, \bibinfo{person}{Sonia
  Castelo}, \bibinfo{person}{Roque Lopez}, \bibinfo{person}{Enrico Bertini},
  \bibinfo{person}{Juliana Freire}, {and} \bibinfo{person}{Claudio Silva}.}
  \bibinfo{year}{2020}\natexlab{}.
\newblock \bibinfo{title}{PipelineProfiler: A Visual Analytics Tool for the
  Exploration of AutoML Pipelines}.
\newblock
\newblock
\showeprint[arxiv]{cs.HC/2005.00160}


\bibitem[\protect\citeauthoryear{Palkar, Thomas, Shanbhag, Narayanan, Pirk,
  Schwarzkopf, Amarasinghe, Zaharia, and InfoLab}{Palkar et~al\mbox{.}}{2017}]%
        {palkar2017weld}
\bibfield{author}{\bibinfo{person}{Shoumik Palkar}, \bibinfo{person}{James~J
  Thomas}, \bibinfo{person}{Anil Shanbhag}, \bibinfo{person}{Deepak Narayanan},
  \bibinfo{person}{Holger Pirk}, \bibinfo{person}{Malte Schwarzkopf},
  \bibinfo{person}{Saman Amarasinghe}, \bibinfo{person}{Matei Zaharia}, {and}
  \bibinfo{person}{Stanford InfoLab}.} \bibinfo{year}{2017}\natexlab{}.
\newblock \showarticletitle{Weld: A common runtime for high performance data
  analytics}. In \bibinfo{booktitle}{\emph{Conference on Innovative Data
  Systems Research (CIDR)}}. \bibinfo{publisher}{CIDR},
  \bibinfo{address}{Chaminade, California}, \bibinfo{pages}{45}.
\newblock


\bibitem[\protect\citeauthoryear{Patel, Bancroft, Drucker, Fogarty, Ko, and
  Landay}{Patel et~al\mbox{.}}{2010}]%
        {Patel2010}
\bibfield{author}{\bibinfo{person}{Kayur Patel}, \bibinfo{person}{Naomi
  Bancroft}, \bibinfo{person}{Steven~M. Drucker}, \bibinfo{person}{James
  Fogarty}, \bibinfo{person}{Andrew~J. Ko}, {and} \bibinfo{person}{James
  Landay}.} \bibinfo{year}{2010}\natexlab{}.
\newblock \showarticletitle{Gestalt: Integrated Support for Implementation and
  Analysis in Machine Learning}. In \bibinfo{booktitle}{\emph{Proceedings of
  the 23nd Annual ACM Symposium on User Interface Software and Technology}}
  \emph{(\bibinfo{series}{UIST '10})}. \bibinfo{publisher}{Association for
  Computing Machinery}, \bibinfo{address}{New York, NY, USA},
  \bibinfo{pages}{37–46}.
\newblock
\showISBNx{9781450302715}
\urldef\tempurl%
\url{https://doi.org/10.1145/1866029.1866038}
\showDOI{\tempurl}


\bibitem[\protect\citeauthoryear{Ramos, Meek, Simard, Suh, and Ghorashi}{Ramos
  et~al\mbox{.}}{2020}]%
        {ramos2020interactive}
\bibfield{author}{\bibinfo{person}{Gonzalo Ramos}, \bibinfo{person}{Christopher
  Meek}, \bibinfo{person}{Patrice Simard}, \bibinfo{person}{Jina Suh}, {and}
  \bibinfo{person}{Soroush Ghorashi}.} \bibinfo{year}{2020}\natexlab{}.
\newblock \showarticletitle{Interactive machine teaching: a human-centered
  approach to building machine-learned models}.
\newblock \bibinfo{journal}{\emph{Human-Computer Interaction}}
  \bibinfo{volume}{35} (\bibinfo{date}{04} \bibinfo{year}{2020}),
  \bibinfo{pages}{1--39}.
\newblock
\urldef\tempurl%
\url{https://doi.org/10.1080/07370024.2020.1734931}
\showDOI{\tempurl}


\bibitem[\protect\citeauthoryear{Roh, Heo, and Whang}{Roh
  et~al\mbox{.}}{2019}]%
        {roh2019survey}
\bibfield{author}{\bibinfo{person}{Yuji Roh}, \bibinfo{person}{Geon Heo}, {and}
  \bibinfo{person}{Steven~Euijong Whang}.} \bibinfo{year}{2019}\natexlab{}.
\newblock \bibinfo{title}{A Survey on Data Collection for Machine Learning: a
  Big Data -- AI Integration Perspective}.
\newblock
\newblock
\showeprint[arxiv]{cs.LG/1811.03402}


\bibitem[\protect\citeauthoryear{Talbot, Lee, Kapoor, and Tan}{Talbot
  et~al\mbox{.}}{2009}]%
        {Talbot2009}
\bibfield{author}{\bibinfo{person}{Justin Talbot}, \bibinfo{person}{Bongshin
  Lee}, \bibinfo{person}{Ashish Kapoor}, {and} \bibinfo{person}{Desney~S.
  Tan}.} \bibinfo{year}{2009}\natexlab{}.
\newblock \showarticletitle{EnsembleMatrix: Interactive Visualization to
  Support Machine Learning with Multiple Classifiers}. In
  \bibinfo{booktitle}{\emph{Proceedings of the SIGCHI Conference on Human
  Factors in Computing Systems}} \emph{(\bibinfo{series}{CHI '09})}.
  \bibinfo{publisher}{Association for Computing Machinery},
  \bibinfo{address}{New York, NY, USA}, \bibinfo{pages}{1283–1292}.
\newblock
\showISBNx{9781605582467}
\urldef\tempurl%
\url{https://doi.org/10.1145/1518701.1518895}
\showDOI{\tempurl}


\bibitem[\protect\citeauthoryear{TransmogriFAI}{TransmogriFAI}{2020}]%
        {TransmogriFAI}
\bibfield{author}{\bibinfo{person}{TransmogriFAI}.}
  \bibinfo{year}{2020}\natexlab{}.
\newblock \bibinfo{title}{TransmogrifAI}.
\newblock \bibinfo{howpublished}{Website}.
\newblock
\urldef\tempurl%
\url{https://github.com/salesforce/TransmogrifAI}
\showURL{%
Retrieved September 5, 2020 from \tempurl}


\bibitem[\protect\citeauthoryear{Vaccaro and Waldo}{Vaccaro and Waldo}{2019}]%
        {AgainstHITL2019}
\bibfield{author}{\bibinfo{person}{Michelle Vaccaro} {and} \bibinfo{person}{Jim
  Waldo}.} \bibinfo{year}{2019}\natexlab{}.
\newblock \showarticletitle{The Effects of Mixing Machine Learning and Human
  Judgment}.
\newblock \bibinfo{journal}{\emph{Commun. ACM}} \bibinfo{volume}{62},
  \bibinfo{number}{11} (\bibinfo{date}{Oct.} \bibinfo{year}{2019}),
  \bibinfo{pages}{104–110}.
\newblock
\showISSN{0001-0782}
\urldef\tempurl%
\url{https://doi.org/10.1145/3359338}
\showDOI{\tempurl}


\bibitem[\protect\citeauthoryear{Van Der~Walt, Colbert, and Varoquaux}{Van
  Der~Walt et~al\mbox{.}}{2011}]%
        {van2011numpy}
\bibfield{author}{\bibinfo{person}{Stefan Van Der~Walt},
  \bibinfo{person}{S~Chris Colbert}, {and} \bibinfo{person}{Gael Varoquaux}.}
  \bibinfo{year}{2011}\natexlab{}.
\newblock \showarticletitle{The NumPy array: a structure for efficient
  numerical computation}.
\newblock \bibinfo{journal}{\emph{Computing in Science \& Engineering}}
  \bibinfo{volume}{13}, \bibinfo{number}{2} (\bibinfo{year}{2011}),
  \bibinfo{pages}{22}.
\newblock


\bibitem[\protect\citeauthoryear{Wang, Weisz, Muller, Ram, Geyer, Dugan,
  Tausczik, Samulowitz, and Gray}{Wang et~al\mbox{.}}{2019b}]%
        {Dakuo2019Collaborative}
\bibfield{author}{\bibinfo{person}{Dakuo Wang}, \bibinfo{person}{Justin~D.
  Weisz}, \bibinfo{person}{Michael Muller}, \bibinfo{person}{Parikshit Ram},
  \bibinfo{person}{Werner Geyer}, \bibinfo{person}{Casey Dugan},
  \bibinfo{person}{Yla Tausczik}, \bibinfo{person}{Horst Samulowitz}, {and}
  \bibinfo{person}{Alexander Gray}.} \bibinfo{year}{2019}\natexlab{b}.
\newblock \showarticletitle{Human-AI Collaboration in Data Science: Exploring
  Data Scientists' Perceptions of Automated AI}.
\newblock \bibinfo{journal}{\emph{Proc. ACM Hum.-Comput. Interact.}}
  \bibinfo{volume}{3}, \bibinfo{number}{CSCW}, Article \bibinfo{articleno}{211}
  (\bibinfo{date}{Nov.} \bibinfo{year}{2019}), \bibinfo{numpages}{24}~pages.
\newblock
\urldef\tempurl%
\url{https://doi.org/10.1145/3359313}
\showDOI{\tempurl}


\bibitem[\protect\citeauthoryear{Wang, Ming, Jin, Shen, Liu, Smith,
  Veeramachaneni, and Qu}{Wang et~al\mbox{.}}{2019a}]%
        {wang2019atmseer}
\bibfield{author}{\bibinfo{person}{Qianwen Wang}, \bibinfo{person}{Yao Ming},
  \bibinfo{person}{Zhihua Jin}, \bibinfo{person}{Qiaomu Shen},
  \bibinfo{person}{Dongyu Liu}, \bibinfo{person}{Micah~J. Smith},
  \bibinfo{person}{Kalyan Veeramachaneni}, {and} \bibinfo{person}{Huamin Qu}.}
  \bibinfo{year}{2019}\natexlab{a}.
\newblock \showarticletitle{ATMSeer: Increasing Transparency and
  Controllability in Automated Machine Learning}. In
  \bibinfo{booktitle}{\emph{Proceedings of the 2019 CHI Conference on Human
  Factors in Computing Systems}} \emph{(\bibinfo{series}{CHI '19})}.
  \bibinfo{publisher}{Association for Computing Machinery},
  \bibinfo{address}{New York, NY, USA}, \bibinfo{pages}{1–12}.
\newblock
\showISBNx{9781450359702}
\urldef\tempurl%
\url{https://doi.org/10.1145/3290605.3300911}
\showDOI{\tempurl}


\bibitem[\protect\citeauthoryear{Weidele, Weisz, Oduor, Muller, Andres, Gray,
  and Wang}{Weidele et~al\mbox{.}}{2019}]%
        {weidele2019autoaiviz}
\bibfield{author}{\bibinfo{person}{Daniel Karl~I. Weidele},
  \bibinfo{person}{Justin~D. Weisz}, \bibinfo{person}{Eno Oduor},
  \bibinfo{person}{Michael Muller}, \bibinfo{person}{Josh Andres},
  \bibinfo{person}{Alexander Gray}, {and} \bibinfo{person}{Dakuo Wang}.}
  \bibinfo{year}{2019}\natexlab{}.
\newblock \bibinfo{title}{AutoAIViz: Opening the Blackbox of Automated
  Artificial Intelligence with Conditional Parallel Coordinates}.
\newblock
\newblock
\showeprint[arxiv]{cs.LG/1912.06723}


\bibitem[\protect\citeauthoryear{{Wikipedia contributors}}{{Wikipedia
  contributors}}{2020}]%
        {wiki:one-hot}
\bibfield{author}{\bibinfo{person}{{Wikipedia contributors}}.}
  \bibinfo{year}{2020}\natexlab{}.
\newblock \bibinfo{title}{One-hot --- {Wikipedia}{,} The Free Encyclopedia}.
\newblock
\newblock
\urldef\tempurl%
\url{https://en.wikipedia.org/w/index.php?title=One-hot&oldid=975049657}
\showURL{%
\tempurl}
\newblock
\shownote{[Online; accessed 17-September-2020].}


\bibitem[\protect\citeauthoryear{Wongsuphasawat, Smilkov, Wexler, Wilson,
  Man{\'{e}}, Fritz, Krishnan, Vi{\'{e}}gas, and Wattenberg}{Wongsuphasawat
  et~al\mbox{.}}{2018}]%
        {Wongsuphasawat2018}
\bibfield{author}{\bibinfo{person}{Kanit Wongsuphasawat},
  \bibinfo{person}{Daniel Smilkov}, \bibinfo{person}{James Wexler},
  \bibinfo{person}{Jimbo Wilson}, \bibinfo{person}{Dandelion Man{\'{e}}},
  \bibinfo{person}{Doug Fritz}, \bibinfo{person}{Dilip Krishnan},
  \bibinfo{person}{Fernanda~B. Vi{\'{e}}gas}, {and} \bibinfo{person}{Martin
  Wattenberg}.} \bibinfo{year}{2018}\natexlab{}.
\newblock \showarticletitle{{Visualizing Dataflow Graphs of Deep Learning
  Models in TensorFlow}}.
\newblock \bibinfo{journal}{\emph{IEEE Transactions on Visualization and
  Computer Graphics}} \bibinfo{volume}{24}, \bibinfo{number}{1}
  (\bibinfo{year}{2018}), \bibinfo{pages}{1--12}.
\newblock
\showISBNx{1077-2626 VO - PP}
\showISSN{10772626}
\urldef\tempurl%
\url{https://doi.org/10.1109/TVCG.2017.2744878}
\showDOI{\tempurl}


\bibitem[\protect\citeauthoryear{Xin, Macke, Ma, Liu, Song, and
  Parameswaran}{Xin et~al\mbox{.}}{2018}]%
        {xin2018helix}
\bibfield{author}{\bibinfo{person}{Doris Xin}, \bibinfo{person}{Stephen Macke},
  \bibinfo{person}{Litian Ma}, \bibinfo{person}{Jialin Liu},
  \bibinfo{person}{Shuchen Song}, {and} \bibinfo{person}{Aditya Parameswaran}.}
  \bibinfo{year}{2018}\natexlab{}.
\newblock \showarticletitle{Helix: holistic optimization for accelerating
  iterative machine learning}.
\newblock \bibinfo{journal}{\emph{Proceedings of the VLDB Endowment}}
  \bibinfo{volume}{12}, \bibinfo{number}{4} (\bibinfo{year}{2018}),
  \bibinfo{pages}{446--460}.
\newblock


\bibitem[\protect\citeauthoryear{Yang, Suh, Chen, and Ramos}{Yang
  et~al\mbox{.}}{2018}]%
        {YangNonExpert}
\bibfield{author}{\bibinfo{person}{Qian Yang}, \bibinfo{person}{Jina Suh},
  \bibinfo{person}{Nan-Chen Chen}, {and} \bibinfo{person}{Gonzalo Ramos}.}
  \bibinfo{year}{2018}\natexlab{}.
\newblock \showarticletitle{Grounding Interactive Machine Learning Tool Design
  in How Non-Experts Actually Build Models}. In
  \bibinfo{booktitle}{\emph{Proceedings of the 2018 Designing Interactive
  Systems Conference}} \emph{(\bibinfo{series}{DIS ’18})}.
  \bibinfo{publisher}{Association for Computing Machinery},
  \bibinfo{address}{New York, NY, USA}, \bibinfo{pages}{573–584}.
\newblock
\showISBNx{9781450351980}
\urldef\tempurl%
\url{https://doi.org/10.1145/3196709.3196729}
\showDOI{\tempurl}


\end{thebibliography}
